\documentclass[a4paper,fleqn,usenatbib]{mnras}
\usepackage[english]{babel} 
\usepackage[T1]{fontenc} 
\usepackage{newtxtext} 
\usepackage[slantedGreek]{newtxmath} 
\hypersetup{pdfauthor={E. P. Lagioia},
            pdftitle={The \textit{Hubble Space Telescope} UV Legacy Survey of Galactic Globular Clusters - XII. The RGB Bumps of multiple stellar populations},
            pdfkeywords={stars: population II, evolution -- globular clusters: general},bookmarksnumbered=true}
\usepackage{graphicx} 
\usepackage{threeparttable}
\usepackage[normalem]{ulem} 
\newcommand{\ngc}{NGC\,}
\newcommand{\hst}{\textit{HST}}
\newcommand{\ome}{$\rm \omega\,Cen$}

\newcommand{\feh}{\rm [Fe/H]}
\newcommand{\afe}{\rm [$\alpha$/Fe]}

\newcommand{\teff}{$T_{\rm eff}$}
\bibpunct[; ]{(}{)}{;}{a}{}{,}
\volume{{\rm in press}} 
\defcitealias{Piotto15}{Paper~I}
\defcitealias{Milone15a}{Paper~II}
\defcitealias{Milone15b}{Paper~III}
\defcitealias{Nardiello15}{Paper~IV}
\defcitealias{Renzini15}{Paper~V}
\defcitealias{Milone17}{Paper~IX}
\defcitealias{Brown16}{Paper~X}

\title[The RGB Bumps of multiple stellar populations]{The \textit{Hubble Space Telescope} UV Legacy Survey of Galactic Globular
Clusters - XII. The RGB Bumps of multiple stellar populations\thanks{Based 
on on observations with the NASA / ESA \hst, obtained at the Space Telescope 
Science Institute, which is operated by AURA, Inc., under NASA contract NAS5-26555.}}

\author[E.~P. Lagioia]{
E.~P. Lagioia,$^{1,2,3}$\thanks{E-mail: elagioia@iac.es}
A.~P. Milone,$^{3,4}$
A.~F. Marino,$^{4}$
S. Cassisi,$^{5}$
A.~J. Aparicio,$^{1,2}$
\newauthor G. Piotto,$^{3,6}$
J. Anderson,$^{7}$
B. Barbuy,$^{8}$
L.~R. Bedin,$^{6}$
A. Bellini,$^{7}$
T. Brown,$^{7}$
\newauthor F. D'Antona,$^{9}$
D. Nardiello,$^{3,6}$
S. Ortolani,$^{3,6}$
A. Pietrinferni,$^{5}$
A. Renzini,$^{6}$
\newauthor M. Salaris,$^{10}$
A. Sarajedini,$^{11}$
R. van der Marel$^{7,12}$
and E. Vesperini$^{13}$
\\
$^{1}$Instituto de Astrof\'{i}sica de Canarias (IAC), calle V\'{i}a L\'{a}ctea s/n, E-38205, La Laguna, Tenerife, Spain\\
$^{2}$Universidad de La Laguna, Dpto. Astrof\'{i}sica, E-38206, Avenida Francisco S\'{a}nchez s/n, E-38206, La Laguna, Tenerife, Spain\\
$^{3}$Dipartimento di Fisica e Astronomia ``Galileo Galilei'', Universit\`{a} di Padova, Vicolo dell'Osservatorio 3, I-35122, Padova, Italy\\
$^{4}$Research School of Astronomy and Astrophysics, The Australian National University, Cotter Road, Weston, ACT, 2611, Australia\\
$^{5}$INAF - Osservatorio Astronomico d'Abruzzo, Via Mentore Maggini, I-64100, Teramo, Italy\\
$^{6}$INAF - Osservatorio Astronomico di Padova, Vicolo dell'Osservatorio 5, I-35122, Padova, Italy\\
$^{7}$Space Telescope Science Institute, 3700 San Martin Dr., Baltimore, MD 21218, USA\\
$^{8}$Universidade de S\~{a}o Paulo, IAG, Rua do Matao 1226, Cidade Universitaria, S\~{a}o Paulo 05508-900, Brazil\\
$^{9}$INAF - Osservatorio Astronomico di Roma, Via Frascati 33, I-00040, Monte Porzio Catone, Roma, Italy\\
$^{10}$Astrophysics Research Institute, Liverpool John Moores University, Liverpool Science Park, IC2 Building, 142 Brownlow Hill, Liverpool L3 5RF, UK\\
$^{11}$Department of Astronomy, University of Florida, 211 Bryant Space Science Center, Gainesville, FL 32611, USA\\
$^{12}$Center for Astrophysical Sciences, Department of Physics \& Astronomy, Johns Hopkins University, Baltimore, MD 21218, USA\\
$^{13}$Department of Astronomy, Indiana University, Bloomington, IN 47401, USA}

\date{Accepted XXX. Received YYY; in original form ZZZ}

\pubyear{2017}

\begin{document}
\label{firstpage}
\pagerange{\pageref{firstpage}--\pageref{lastpage}}
\maketitle

\begin{abstract}
The {\it Hubble Space Telescope} UV Legacy Survey of Galactic Globular Clusters
is providing a major breakthrough in our knowledge of Globular Clusters (GCs)
and their stellar populations. Among the main results, we discovered that all
the studied GCs host two main discrete groups consisting of first generation
(1G) and second generation (2G) stars.
  
We exploit the multiwavelength photometry from this project to investigate, for
the first time, the Red Giant Branch Bump (RGBB) of the two generations in a
large sample of GCs. We identified, with high statistical significance, the RGBB
of 1G and 2G stars in 26 GCs and found that their magnitude separation as a
function of the filter wavelength follows comparable trends. 

The comparison of observations to synthetic spectra reveals that the RGBB
luminosity depends on the stellar chemical composition and that the 2G RGBB is
consistent with stars enhanced in He and N and depleted in C and O with respect
to 1G stars. 

For metal-poor GCs the 1G and 2G RGBB relative luminosity in optical bands
mostly depends on helium content, Y. We used the RGBB observations in F606W and
F814W bands to infer the relative helium abundance of 1G and 2G stars in 18 GCs,
finding an average helium enhancement $\rm \Delta Y = 0.011\pm0.002$ of 2G
stars with respect to 1G stars.
 
This is the first determination of the average difference in helium
abundance of multiple populations in a large number of clusters and provides a
lower limit to the maximum internal variation of helium in GCs.

\end{abstract}


\section{Introduction}\label{sec:int}   
The discovery of multiple stellar sequences in the photometric diagrams of
Globular Clusters (GCs) in our Galaxy has raised, in the last years, new questions
about the physical and dynamical processes responsible for the chemical
inhomogeneities observed in different generations of stars 
(\citealt{Bedin04,Piotto15}, hereafter Paper~I; \citealt{Milone17}, hereafter Paper~IX).

Indeed, several photometric and spectroscopic studies have demonstrated that
the split of the characteristic evolutionary sequences in colour magnitude
diagrams (CMDs) is connected to the variation, in the surface abundance of
stars, of elements produced through CNO cycling and hot proton-capture processes
(C-N, Na-O, Mg-Al anti-correlations) and, in turn, to helium content differences 
among different stellar populations \citep[see][and references therein]{Gratton12}.

Various theories have been put forward so far about the nature of the primordial
generation of polluters, including Asymptotic Giant Branch (AGB) stars
\citep{DAntona05,Dercole10,DAntona16}, fast rotating massive stars
\citep{Decressin07}, massive interacting binaries \citep{deMink09} and
supermassive stars \citep{Denissenkov14}, but no firm conclusion has still been
achieved \citep[see][hereafter Paper~V, for a critical confrontation of these
scenarios with the observational constraints]{Renzini15}.

Each proposed scenario envisages different yields of helium in the enriched
material out of which second-generations stars are formed but we still miss
information on the helium content of the different stellar populations in most
GCs.

Direct measurements based on the of helium lines in the spectra of GC stars are,
in fact, limited to Horizontal Branch (HB) stars with effective temperature
(\teff) in the range $\sim 8,000$ -- $11,500$ K
\citep{Villanova09,Marino14},
where the cold boundary corresponds to the \teff\ at which appear the first
optical \ion{He}{I} lines and the hot boundary to the \teff\ at which
atomic diffusion processes (gravitational settling) begin to significantly alter
the surface helium abundance \citep[][hereafter Paper~X]{Grundahl99,Brown16}. 

An alternative method based on the observation of the chromospheric
near-infrared \ion{He}{I} transition line at $10\,830$ \AA\ in Red Giant
Branch (RGB) stars has been employed by \citet{Dupree11} and \citet{Pasquini11}
for the clusters \ngc5139 (\ome) and \ngc2808, although for a few RGB stars only.

On the other hand, a multiwavelength photometric approach based on the
comparison of the observed colour difference of the multiple Main Sequences (MSs)
and/or RGBs with appropriate theoretical models has been applied to few GCs,
namely \ngc104, \ngc288, \ngc2419, \ngc2808, \ngc5139, \ngc6352, \ngc6397,
\ngc6441, \ngc6752, \ngc7078 and \ngc7089, for which the relative
helium abundance of the different sub-populations was estimated
(\citealt{Piotto05,Piotto07,Villanova07,diCriscienzo11,Milone12b,Milone12c,Bellini13,Milone13,Milone14};
\citealt{Nardiello15}, hereafter Paper~IV).

Alongside the colour difference, an important evolutionary feature
sensitive to the helium content of an old stellar population is the RGB Bump
(RGBB) that, in a CMD, appears as a clump of stars along the RGB
\citep[e.g.][and references therein]{Cassisi97}. It is as well recognizable as an
excess in the histogram distribution of the magnitude, or differential
luminosity function (LF), of the RGB stars \citep{Thomas67,Iben68}. The RGBB is
produced by the three-fold passage of the RGB stars through the same luminosity
interval during their evolution.  

Indeed, during the ascent of the RGB, the hydrogen-burning shell of a star
steadily moves outward thereby approaching the chemical discontinuity left
behind by the first dredge up. The increase in the opacity due to the larger
hydrogen abundance just above the shell causes a temporary drop in the stellar
luminosity that, once the shell has gone through the discontinuity, starts again
to grow monotonically \citep{Sweigart90}.  

While the amplitude of the chemical discontinuity affects the RGBB lifetime,
which is reflected in the size (peak height) and shape (width) of the LF excess
\citep{Bono01,Nataf14}, its characteristic brightness is determined by the
maximum penetration of the convective envelope that, in turn, depends on
parameters like age, metal abundance and helium content of the stars
\citep{Cassisi16}. For these reasons the RGBB shape and luminosity represent
fundamental tools to probe the inner chemical profile of the Red Giant stars and
constraints on the knowledge of the aforementioned stellar parameters in a
cluster.
In particular, an increase of helium makes, at fixed age and metallicity, the
RGBB luminosity brighter (and the lifetime shorter). As a consequence, the RGBB
location can be used to constrain the relative helium abundance of multiple
populations in GCs. However, a RGBB magnitude separation caused by variations of
a few percent in mass fraction of helium content, Y, in
different stellar populations, is only detectable with homogeneous, high
precision photometry. 

An empirical investigation on the correlation between RGBB magnitude difference
and helium content variation was presented in \citet{Bragaglia10}, who measured
the displacement in V band of the LF peak of a combined sample of 1368 Na-poor
and Na-rich Red Giants belonging to 14 GCs, corresponding to an average
abundance difference in Y of $0.01\pm0.01$ by assuming the
same heavy elements distribution in the Na-poor and Na-rich stars. Similarly,
\citet{Nataf11} concluded that the gradient of the RGBB brightness and star
counts with the radial distance observed in \ngc104 (47\,Tuc), was consistent
with the presence of a helium enriched stellar population in the cluster centre. 

A previous attempt to use the RGBB magnitude difference to infer helium content
variations in a GC has been done by \citet[][hereafter Paper~III]{Milone15b} for
\ngc2808, in which two out of the five stellar populations of the cluster have
a RGBB which location is compatible with a helium enrichment of $\rm Y\sim0.03$ and
$\sim 0.10$ with respect to the cluster reference population.

In this paper and in \citetalias{Milone15b} we assumed that the reference
population has standard helium abundance $\rm Y = 0.245+1.4\,Z$, where Z is the
cluster metallicity \citep{Pietrinferni04}.  Hence, we assumed in
\citetalias{Milone15b} an helium abundance of $\rm Y = 0.248$ for the primordial
stellar component of \ngc2808. We also verified that the resulting helium
enhancement inferred for the second generation does not depend on assumed helium
abundance of 1G stars.

The aim of this work, which is part of the \textit{Hubble Space Telescope}
(\hst) UV Legacy Survey of Galactic Globular Clusters \citepalias{Piotto15}, is
to identify, for the first time, the RGBB of the distinct stellar populations of
56 GCs and constrain their differential content in helium. 

Our sample includes \ngc2808, which has been analysed in \citetalias{Milone15b}
using a similar method. We have excluded \ngc5139 because the multiple stellar
populations in this cluster show an extreme degree of complexity
\citep[e.g.][]{Johnson09,Marino11,Bellini17} that deserves a separate analysis.    

The paper is organized as follows. In Section~\ref{sec:obs} we describe the
photometric catalogues, the data reduction techniques and the selection criteria
of the stellar populations. The methods used to the determine the luminosity of
the RGB bump for the distinct stellar populations in each GCs are described in
Section~\ref{sec:RGBLF}. The comparison with theoretical models and the helium
abundance estimates are illustrated in Section~\ref{sec:deltaY}. Finally,
Section~\ref{sec:end} provides the summary and the discussion of the results.

\section{Cluster database and data reduction}\label{sec:obs}   
In the present work we have used the photometric catalogues, presented in
\citetalias{Piotto15} and \citetalias{Milone17}, of the \hst\ UV Legacy Survey
of GCs program (G0-13297, PI G. Piotto), in UV (F275W and F336W) and optical
(F438W) bands, obtained with the Ultraviolet and Visual Channel of the Wide
Field Camera 3 (WFC3/UVIS) on board \hst. These observations were complemented
with WFC3/UVIS data collected in the same filters for GO-12605 and GO-12311 (PI
G. Piotto) and from archive data that, in fact, extend the spectral coverage of
the ACS/WFC optical (F606W and F814W) data of the clusters observed in the ACS
Survey of GCs Treasury Program (G0-10775, PI A.  Sarajedini).

Since the details on the entire dataset, exposure times and data reduction have
been already provided in \citetalias{Piotto15} and \citetalias{Milone17}, here
we briefly summarize the data reduction technique employed to obtain the final
catalogues.  

Each individual UVIS exposure has been corrected for poor charge-transfer
efficiency according to the solution provided by \citet{Anderson10}. The
photometric reduction has been performed with the software
\texttt{img2xym\_wfc3uv}, developed by Jay Anderson and mostly based on the
program \texttt{img2xym\_WFC} \citep{AndKing06}. The magnitude of saturated
stars was recovered from saturated images with the procedure described in
\citet{Anderson08}, developed from the method of \citet{Gilliland04}, which
takes into account the electrons bled into adjacent pixels. Stellar positions
have been corrected for geometric distortion according to the solution of
\citet{Bellini11}. Calibration to the VEGAMAG system was performed by applying
the encircled energy distribution and zero points listed in the STScI website
for the UVIS
detector\footnote{\url{http://www.stsci.edu/hst/acs/analysis/zeropoints/zpt.py}},
following the recipe of \citet{Bedin05}. ACS/WFC data reduction has already been
described in \cite{Anderson08} and we refer the interested reader to this paper
for details about the adopted procedures.

The catalogues have been purged from non-cluster members and photometric outliers,
taking into account cluster proper motions and the quality indexes provided by
the reduction software. Proper motions have been obtained by comparing the
average stellar positions in two epochs \citep{Anderson03,Piotto12}, derived
from WFC3 F336W/F438W images and ACS catalogues by \citet{Anderson08}. Quality
indexes provided by the software \citep{Anderson06,Anderson08} allowed to select
stars measured with high-accuracy in the proper-motion selected catalogues, by
adopting the procedure detailed in \citet{Milone09}. Finally the correction for
differential reddening and Point Spread Function spatial variation, illustrated
in \citet{Milone12a} was applied to each catalogue.

\subsection{Selection of first and second stellar populations}\label{subsec:sel}
Spectroscopic investigation on the chemical signature of multiple populations in
GCs has shown that the second stellar generations\footnote{The expressions
stellar ``population'' and ``generation'' are used as synonyms in this work.}
are enriched in sodium and nitrogen and depleted in carbon and oxygen, thus
indicating that they formed from material exposed to some degree of CNO
processing \citep{Prantzos06}.

In this context, the F275W, F336W and F438W filters are ideal tools for the
photometric identification of the distinct stellar populations because their
passband encompasses the absorption wavelengths of the OH, NH, CN and CH
molecules, respectively \citepalias{Milone15b}. 

This is clearly demonstrated in the $\rm m_{F336W}$ vs.\,$\rm
C_{F275W,F336W,F438W}$ diagrams displayed in \citetalias{Piotto15}. Indeed the
pseudo-colour $\rm C_{F275W,F336W,F438W} =
(m_{F275W}-m_{F336W})-(m_{F336W}-m_{F438W})$, is an efficient tool to separate
the multiple stellar populations along the MS, RGB and AGB. 

In addition, the wide colour baseline provided by the F275W and F814W magnitudes
considerably improves the sensitivity of our observations to the stellar
temperature and, in turn, to helium abundance variation among multiple
populations in a cluster. 

To identify the two main populations we used the $\rm \Delta_{C\ F275W,F336W,F438W}$
vs.\,$\rm \Delta_{F275W,F814W}$ diagrams, introduced and discussed in
\citetalias{Milone15b} and \citetalias{Milone17}. These diagrams
have been referred to as chromosome maps in \citetalias{Renzini15} and we will
keep using this denomination in the following. The quantities $\rm
\Delta_{F275W,F814W}$ and $\rm \Delta_{C\ F275W,F336W,F438W}$ are indicative of
the relative distance of a star with respect to the blue and red boundary of the
RGB $\rm (m_{F275W}-m_{F814W})$ colour and the $\rm C_{F275W,F336W,F438W}$
pseudo-colour, respectively.

\begin{figure*} 
\includegraphics[width=0.87\textwidth]{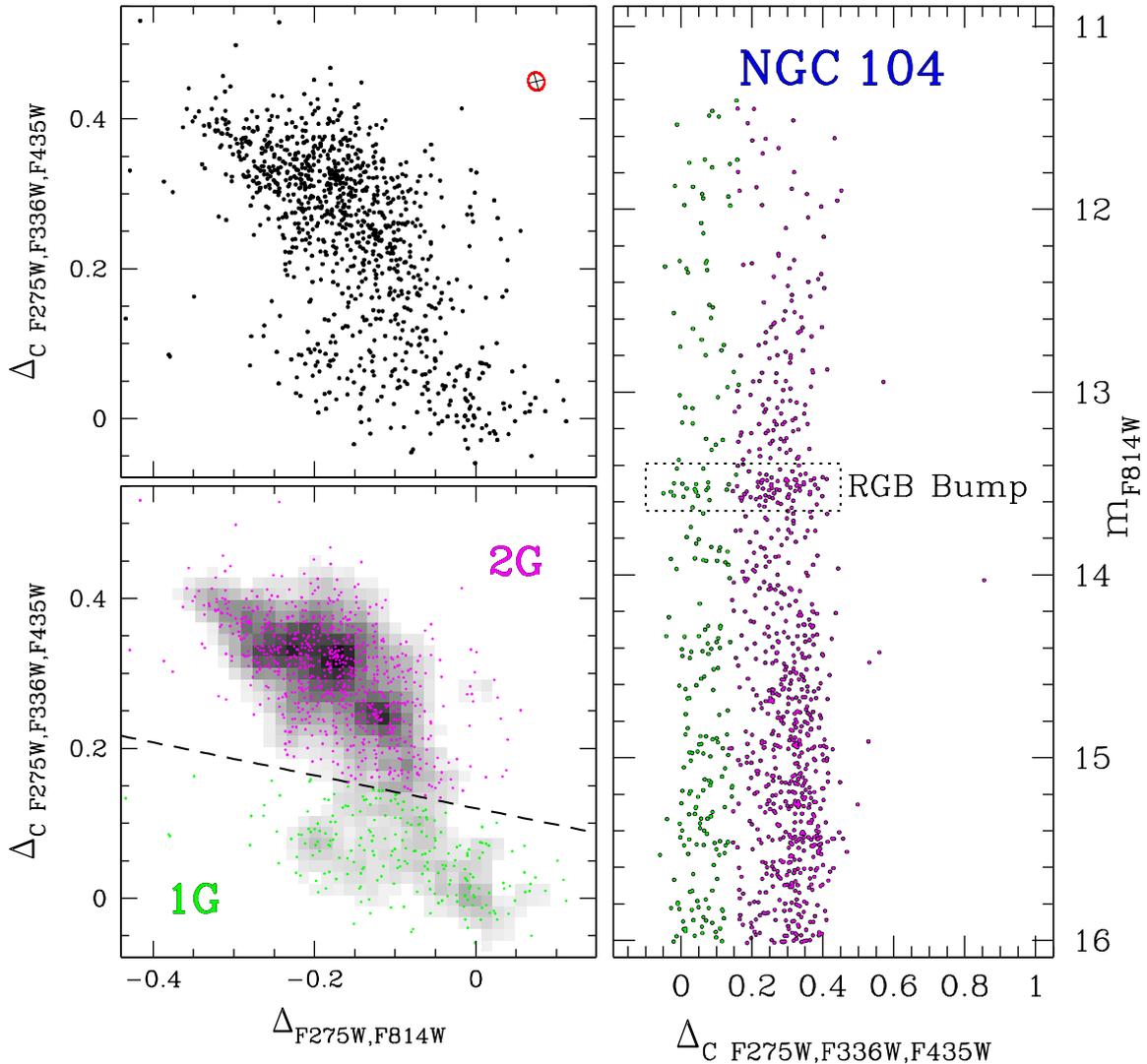}
\caption{Upper-left panel: $\rm \Delta_{C\ F275W,F336W,F435W}$ vs.\,$\rm
\Delta_{F275W,F814W}$ diagram (chromosome map) of the RGB stars of the GC
\ngc104. The red ellipse marks the 68.27th percentile of the expected
distribution of the observational errors in this diagram
\citepalias[see][]{Milone17}. Lower-left panel: stellar density diagram relative
to the chromosome map. The black dashed line divides the first (1G) and second
(2G) stellar population in the chromosome map. 1G and 2G stars have been plotted
over the density diagram as green and magenta dots, respectively. Right panel:
$\rm m_{F814W}$ vs.\,$\rm \Delta_{C\ F275W,F336W,F435W}$ diagram of the cluster
RGB.  The dotted box indicates the approximate position of the 1G and
2G RGB Bump stars.\label{fig:chm}}
\end{figure*}

For example, in the upper-left panel of Figure~\ref{fig:chm} we show the
chromosome map of \ngc104\footnote{For this cluster we used images collected
through the F435W filter of the ACS/WFC, which is very similar to the F438W
filter of WFC3/UVIS used for most GCs. The difference between photometry in the
two filters is negligible for our purposes (see \citetalias{Piotto15} for
details).}. A glance at the plot indicates that the star distribution is far
from being homogeneous. This impression is confirmed by the corresponding
stellar density diagram, shown in the lower-left panel, which displays a group
of prominent clumps elongated from the upper-left to the lower-right corner and
centred at $\rm (\Delta_{F275W,F814W}, \Delta_{C\
F275W,F336W,F435W})\sim(-0.17,-0.32)$ and other minor clumps aligned in an
almost horizontal band at $\rm \Delta_{C\ F275W,F336W,F435W} \lesssim 0.15$. In
order to tag the two main stellar populations visible in this diagram, we took
advantage of the selection already performed in \citetalias{Milone17} for all
the GCs observed in the UV Legacy Survey of Galactic Globular Clusters. The plot
is divided into two parts by a black dashed line. All the stars located below
the line are tagged as first generation(s) (1G) and plotted over the density
diagram as green dots, while the stars located above the line are tagged as
secondary generation(s) (2G) and plotted as magenta dots. In the rest of the
paper we will use the same colour code for 1G and 2G stars. 

We note in passing that substructures are visible for each stellar group in the
density diagram thus demonstrating that both 1G and 2G stars host stellar
sub-populations (see for instance \citealt{Milone15a}, hereafter Paper~II, and
\citetalias{Milone15b} and \citetalias{Milone17}). In this paper we are
interested in the average properties of 1G and 2G stars and we focus on these
two main populations only.

\citetalias{Milone17} has shown that the chromosome maps of a sample of
GCs, including \ngc362, \ngc1261, \ngc1851, \ngc5139, \ngc5286, \ngc6388,
\ngc6656, \ngc6715, \ngc6934, \ngc7089 show a split of both 1G and 2G sequences
and a split SGB, which is also visible in optical CMDs. The faint and the bright
SGBs are connected to the blue and red RGBs, respectively, in the $m_{\rm
F336W}$ vs.\,$m_{\rm F336W}-m_{\rm F814W}$ CMD \citep[see
also][]{Han09,Marino11,Marino15}. Spectroscopy reveals that red-RGB stars are
enhanced in s-process elements, iron, and overall $\rm C+N+O$ abundance with
respect to blue-RGB stars
\citep{Marino09,Marino12,Marino15,Yong08,Yong09,Yong14,Carretta10,Johnson15,Johnson17}.
The relative luminosity of the RGB bump of the blue and the red RGB is affected
by their relative abundance of iron and $\rm C+N+O$. Since these quantities are
poorly known in most of the analyzed clusters, in this paper we focus on
blue-RGB stars only.

In the right panel of the figure we plotted the $\rm m_{F814W}$ vs.\,$\rm
\Delta_{C\ F275W,F336W,F435W}$ diagram of the RGB stars of the cluster. We also
highlighted, with a black box, the approximate location of the 1G and 2G RGBBs.
In the next two sections we will describe in detail the method used to determine
the RGBB luminosity and infer the average difference in helium abundance between
the 1G and 2G in each cluster.

\section{Determination of the RGB bumps luminosity}\label{sec:RGBLF}   

\begin{figure*}  
\includegraphics[width=0.85\textwidth]{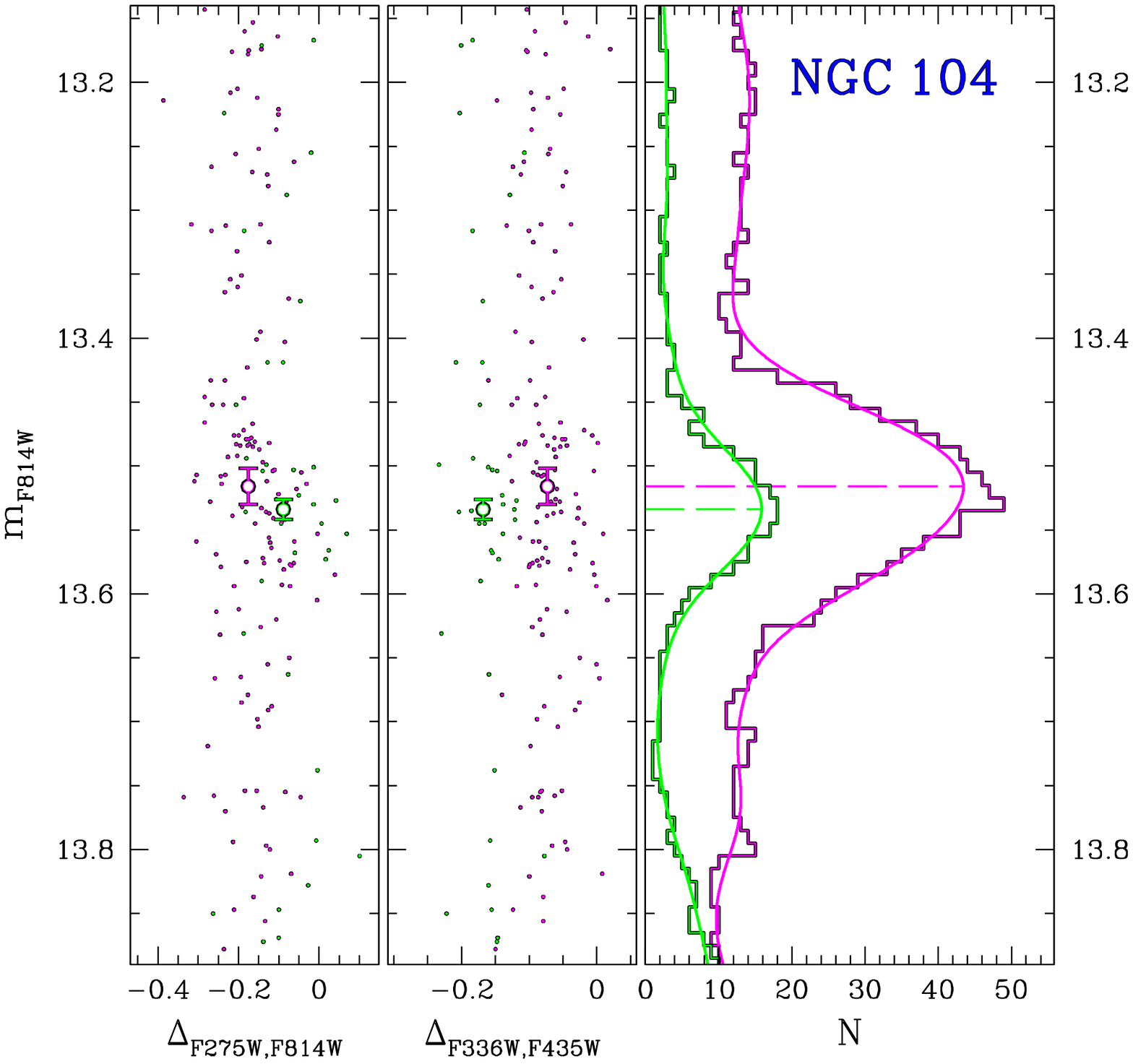}
\caption{Left and middle panel: $\rm m_{F814W}$ vs.\,$\rm \Delta_{F275W,F814W}$ (left)
and $\rm m_{F814W}$ vs.\,$\rm \Delta_{F336W,F435W}$ (right) diagram of the RGB of the
GC \ngc104, approximately centred on the 1G and 2G RGBBs, marked by the green
and magenta open circles, respectively. Right panel: F814W LF of the 1G (green
histogram) and 2G (magenta histogram) stars displayed in the left and middle
panel. The continuous line superimposed on each LF represents the corresponding
kernel density estimate, which maximum indicates the RGB Bump magnitude. The
uncertainty associated to the bump magnitude is plotted as a vertical error bar
for the green and magenta open circles in the left and middle panels.\label{fig:histo}} 
\end{figure*}

In this section we describe the method to estimate the magnitude difference
between the RGBB of 1G and 2G stars in all the studied GCs. As an example, we
illustrate in Figure~\ref{fig:histo} the procedure, in the F814W band, for the
GC \ngc104.

The $\rm m_{F814W}$ vs.\,$\rm \Delta_{F275W,F814W}$ diagram of a portion of the
cluster RGB is shown in the left panel of Figure~\ref{fig:histo}. The location
of the 1G and 2G RGBB is indicated by the green and the magenta bold point,
respectively. For the sake of comparison, the same stars as well as the two
bumps are plotted in the $\rm m_{F814W}$ vs.\,$\rm \Delta_{F336W,F435W}$
diagram, displayed in the middle panel, where the quantity $\rm
\Delta_{F336W,F435W}$ is the analogous of $\rm \Delta_{F275W,F814W}$ but in the
$\rm m_{F814W}$ vs.\,$\rm (m_{F336W}-m_{F435W})$ diagram.

To determine the position of the RGBB of each population we first defined a
magnitude bin $w=0.1$ mag, and then built the LF of the 1G and 2G stars
over a range of 0.8 mag centred on the approximate location of their RGBB, as
shown in the right panel of Figure~\ref{fig:histo}. Both the LFs were obtained
by adopting the method of the naive estimator of \citet{Silver86}. In a
nutshell, we divided the analysed magnitude range into a regular grid of ${\rm
m}_{\rm F814W}^{i}$ points $w/10$ mag apart and, for each point, we
counted the number of stars in the interval ${\rm m}_{\rm F814W}^{i}-w/2 < {\rm
m_{F814W}} < {\rm m}_{\rm F814W}^{i}+w/2$.

To determine the RGBB luminosity, $\rm m_{F814W,bump}$, of both the 1G and 2G
stars we constructed, for each LF, a kernel density estimator by using a
Gaussian kernel function with variance equal to $4/25\,w^2$.

The resulting kernel density estimates of the 1G and 2G stars are plotted,
respectively, as green and magenta solid lines in the right panel of
Figure~\ref{fig:histo}. The magnitude corresponding to the maximum of each
function has been taken as our best estimate of the bump luminosity for that
population in the F814W band.

The error associated with each $\rm m_{F814W,bump}$ estimate was computed by
carrying out 1,000 bootstrapping tests on random sampling with replacement of
the RGB stars in the selected magnitude interval. The 68.27th percentile of the
distribution of the bootstrapped $\rm m_{F814W,bump}$ measurements was
considered the standard error of the RGBB magnitude estimate. The vertical error
bar associated to the green and magenta bold points in the left and middle
diagram represents, respectively, the 1G and 2G $\rm m_{F814W,bump}$ uncertainty
for the cluster \ngc104.

We applied the above procedure to derive the RGBB magnitude and the
corresponding uncertainty also in the F275W, F336W, F438W and F606W bands.

\begin{figure}
\includegraphics[width=\columnwidth]{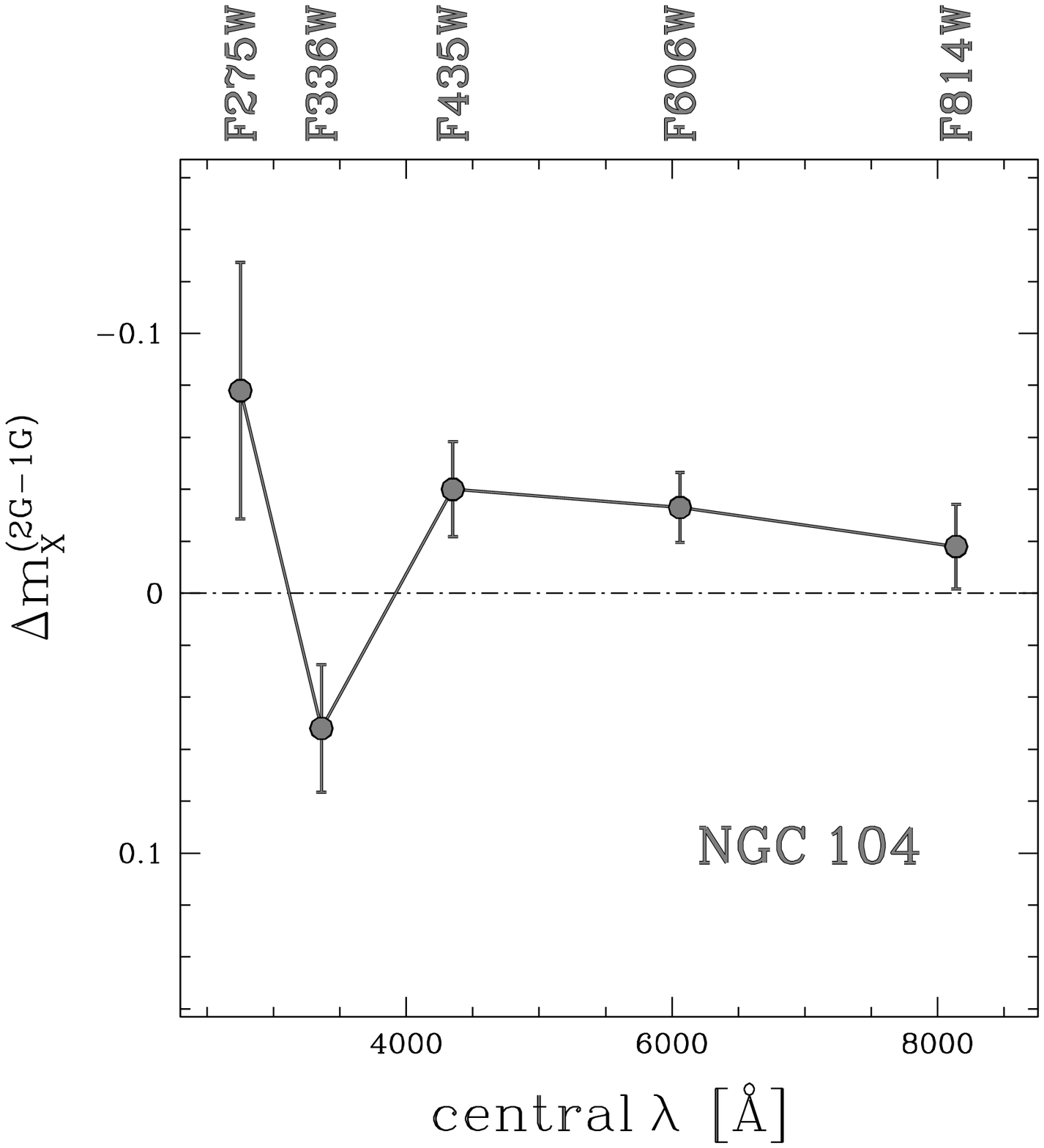}
\caption{Magnitude difference between the 2G and 1G RGB bump of \ngc104
vs.\,central wavelength of the F275W, F336W, F435W, F606W, F814W
filters.\label{fig:delta}}
\end{figure}

Figure~\ref{fig:delta} shows the magnitude difference between the RGBB of the 2G
and 1G of \ngc104, $\rm \Delta m_{X}^{(2G,1G)}$, as a function of the central
wavelength of the filter X, with X = F275W, F336W, F435W, F606W, F814W. The grey
dots indicate the observed difference and the corresponding error bars have been
obtained by adding in quadrature the error of the 1G and 2G RGBB magnitude.  We
find that $\rm \Delta m_{X}^{(2G,1G)}$ is negative in the F275W, F435W, F606W
and F814W bands but positive in the F336W band.

This outcome derives from the combination of the adopted filters and the
chemical properties of the two different cluster populations. Indeed, the F275W
and F435W (or F438W for WFC3/UVIS) filter passband encompasses, respectively,
OH bands and CN and CH bands, while the F336W NH bands. Since 1G stars
are carbon- and oxygen-rich and nitrogen-poor, they appear brighter than 2G
stars in the F336W band and fainter in the F275W and F435W bands. Conversely 2G
stars, formed by CNO-processed material, are carbon- and oxygen-poor and
nitrogen-rich, therefore appearing fainter than 1G stars in F336W band and
brighter in the F275W and F435W (F438W) bands \citepalias{Piotto15}. The optical
F606W and F814W filters are instead mostly sensitive to the stellar \teff, thus
indicating that 2G stars are on average hotter than 1G stars, possibly due to
their enhanced helium content \citep{Milone12b}.

\subsection{Statistical significance of the RGB Bumps}
\begin{figure*}
\includegraphics[width=0.85\textwidth]{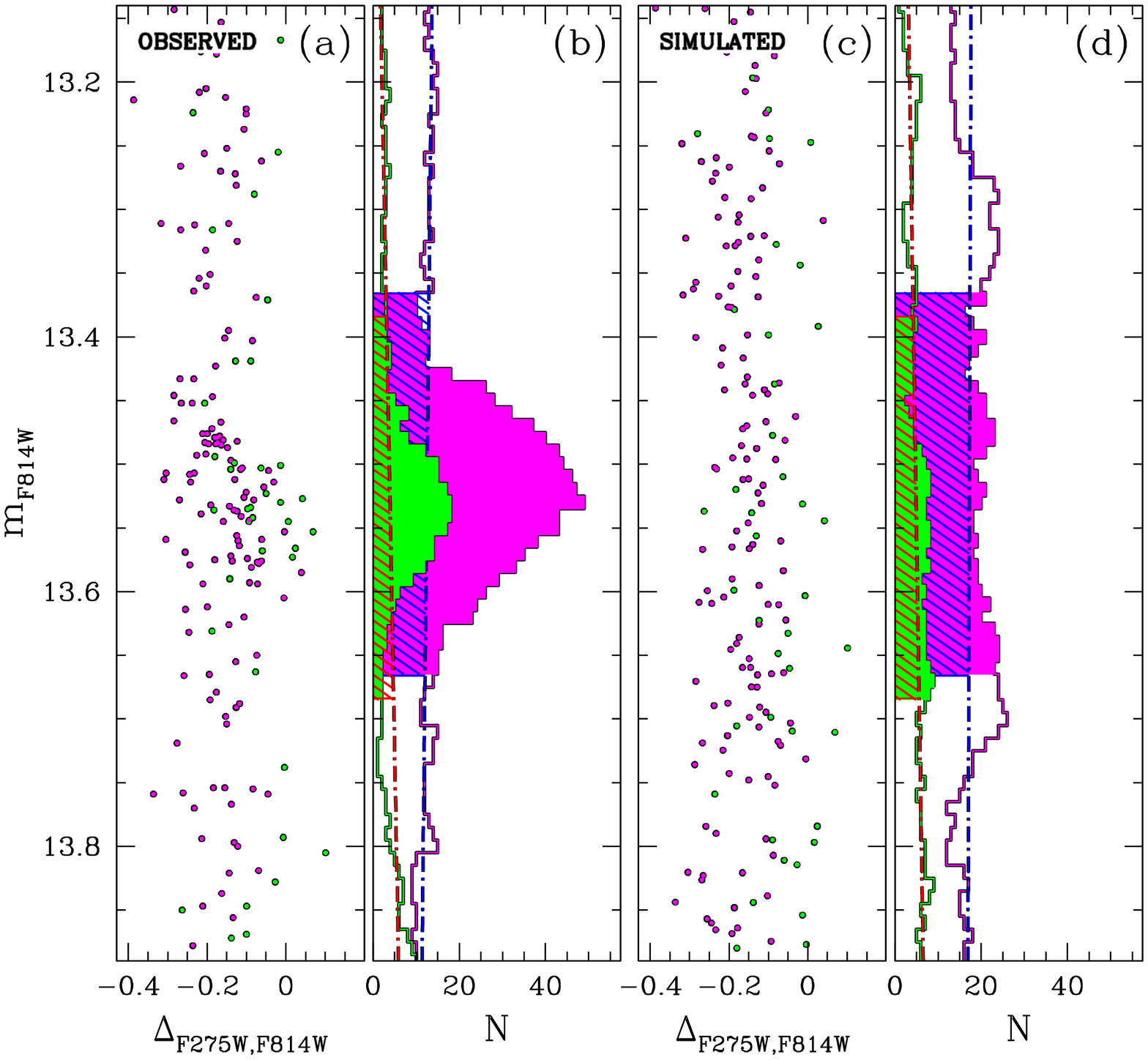}
\caption{Estimate of the significance of the observed RGB Bump for the two
stellar populations in \ngc104. In panel (a) is shown the $\rm m_{F814W}$
vs.\,$\rm \Delta_{F275W,F814W}$ diagram of the 1G (green) and 2G (magenta) RGB
stars of the cluster while the relative LFs are plotted in panel (b). An example
of simulated $\rm m_{F814W}$ vs.\,$\rm \Delta_{F275W,F814W}$ diagram is shown in
panel (c), whereas the corresponding 1G and 2G synthetic LFs are plotted in
panel (d). In panel (b) and (d), the red and blue dot-dashed lines represent,
respectively, the best-fit lines of the 1G and 2G LF. The shadowed portions of
each histogram correspond to the RGBB segments (see text for
details).\label{fig:signif}} 
\end{figure*}

The statistical significance of the RGBB detection is inferred by comparing the
observations with a sample of 10,000 simulated $\rm m_{F814W}$ vs.\,$\rm
\Delta_{F275W,F814W}$ diagrams. This method, similar to that used in
\citetalias{Milone15b}, is described in the following and illustrated, for the
GC \ngc104, in Figure~\ref{fig:signif}.  In panel (a) of Figure~\ref{fig:signif}
we show the $\rm m_{F814W}$ vs.\,$\rm \Delta_{F275W,F814W}$ diagram of the two
stellar populations of the cluster while in panel (b) we show corresponding LFs.
The red and blue dot-dashed lines plotted over the LFs are the lines of best fit
of the 1G and 2G LFs, respectively. Since the computation of the best-fit line
of a RGB LF could be affected by the RGBB overdensity, we decided to exclude,
for each LF, all the points within 0.15 mag from the corresponding RGBB. The
excluded portions of the 1G and 2G LFs have been coloured green and magenta,
respectively. For simplicity, in the following we indicate the corresponding
magnitude interval as RGBB segment, while the difference between the area of
each LF and the area below the corresponding best-fit line, in the RGBB segment,
as $\rm dA_{obs}$. The fact that $\rm dA_{obs}$ is greater than zero can be
either an intrinsic feature of the LF itself due to the RGBB or can be an
artefact due to the photometric errors and the small number of analysed stars. 

To discriminate between these two possibilities, we applied the following method
to the 1G and 2G stars separately. For each stellar population we simulated
10,000 $\rm m_{F814W}$ vs.\,$\rm \Delta_{F275W,F814W}$ diagrams, each composed
of 5,000 artificial stars. By construction, the LF of each simulated diagram has
the same input slope of the observed LF best-fit line. Hence a sub-sample with
the same number of stars as in the observed diagram was randomly extracted from
the simulated stars. For each extracted sample of stars, we derived the LF$_i$
and its slope by following the same prescriptions used for the observations.
Then, we calculated the difference ${\rm dA}_{i,\rm sim}$ between the area of
the LF in the RGBB segment and the area below the corresponding best-fit line.
If ${\rm dA}_{i,\rm sim}$ were systematically smaller than $\rm dA_{obs}$, then
the observed stellar overdensity would be likely due to the presence of a RGBB.
In the opposite case, the observed stellar overdensity would be likely
associated to a fluctuation of the LF due to the small number of analysed stars.
The statistical significance is thus defined as the percentage of times on the
total number of simulations in which the relation ${\rm dA}_{i,\rm sim} < \rm
dA_{obs}$ is satisfied. An example of simulated diagram is provided in panel (c)
of Figure~\ref{fig:signif}, while the corresponding LF is shown in panel (d).

In the case of \ngc104 we find that the RGBB is significant at the 99.6\% level
for the 1G stars and 100\% for the 2G stars, thus demonstrating that the
observed stellar overdensities observed along the two main RGBs of this cluster
are due to the presence of the corresponding RGBBs.

\section{Characterization of the RGB bumps}\label{sec:deltaY}   
The procedure described in the previous section for \ngc104 has been
applied to all the 56 GCs analyzed in this work, for which we derived the 1G and
2G RGBB magnitude and the corresponding error in all the filters as well as the
1G and 2G RGBB statistical significance in the F606W and F814W bands. For each
GC, the LF was obtained by adopting $w=0.1$ mag if at least 40 stars in
either the 1G or 2G sample were present in the selected 0.8 mag interval
centered around the approximate location of the cluster 1G and 2G RGBBs, in the
F814W band; if not $w=0.2$ mag was adopted. In both cases a grid step
of $w/10$ mag was used. 

Because of the poor statistics we decided to exclude from further analysis the
19 clusters with less than 15 stars in both the 1G and 2G sample in the selected
F814W magnitude range, namely \ngc288, \ngc2298, \ngc3201, \ngc4590, \ngc5053,
\ngc5466, \ngc5897, \ngc6101, \ngc6121, \ngc6144, \ngc6218, \ngc6366, \ngc6397,
\ngc6535, \ngc6717, \ngc6779, \ngc6809, \ngc6838 and \ngc7099. Indeed, for these
clusters it is not possible to unambiguously identify the presence of the 1G or 2G
or both RGBBs. 

In twelve clusters, namely \ngc1261, \ngc5024, \ngc5286, \ngc5904, \ngc6254,
\ngc6341, \ngc6496, \ngc6541, \ngc6584, \ngc6637, \ngc6656 and \ngc7089, the
significance of the RGBB of at least one population is smaller than 90\% in the
F814W band. Therefore, because of the low significance of their RGBB, we decided
to exclude from the following analysis all the previous clusters except \ngc5904
and \ngc6637, both with a 1G RGBB significance marginally below the adopted 90\%
threshold. Finally, we excluded \ngc2808 because the RGBB of the multiple
populations in this cluster was already analysed in \citetalias{Milone15b}.

Table~\ref{tab:tab1} gives the list of the 26 clusters with 1G and 2G
RGBB-detection significance $\gtrsim 90\%$, for which we indicate the magnitude
difference between the 2G and 1G RGBB, $\rm \Delta m_{X}^{(2G,1G)}$, in all the
bands.

\begin{table*}
\centering
\caption{List of the 26 selected GCs with 1G and 2G RGB-Bump-detection significance 
$\gtrsim 90\%$. The columns 2--6 give the
observed difference between the RGB Bump magnitude of the 2G and 1G stars in the
F275W, F336W, F438W, F606W, F814W filter, respectively. Columns 7--8 and 9--10
give the statistical significance of the 1G--2G RGB Bumps in the F606W and F814W
band, respectively.\label{tab:tab1}}
\begin{threeparttable}
\normalsize
\begin{tabular}{lr@{$\,\pm\,$}lr@{$\,\pm\,$}lr@{$\,\pm\,$}lr@{$\,\pm\,$}lr@{$\,\pm\,$}lrrrr}
\hline
         & \multicolumn{10}{c}{\ \ $\rm \Delta m_{X}^{(2G,1G)}$ (mag)}  & \multicolumn{4}{c}{\ \ Significance (\%)} \\ 
 Cluster & \multicolumn{2}{c}{F275W} & \multicolumn{2}{c}{F336W} & \multicolumn{2}{c}{F438W} & \multicolumn{2}{c}{F606W} & \multicolumn{2}{c}{F814W} & \multicolumn{2}{c}{\ \ F606W} & \multicolumn{2}{c}{\ \ F814W} \\
 & \multicolumn{10}{c}{} & 1G\ \ \ & 2G\ \ \ & 1G\ \ \ & 2G\ \ \ \\
\hline
 \ngc104\tnote{*}          & -0.078&0.049  &  0.052&0.025 & -0.040&0.018 & -0.033&0.013 & -0.018&0.016 &  99.0  & 100.0  &  99.6  & 100.0 \\
 \ngc362                   & -0.079&0.016  &  0.044&0.026 & -0.023&0.018 & -0.023&0.013 & -0.024&0.012 &  94.7  & 100.0  &  94.0  & 100.0 \\
 \ngc1851                  & -0.080&0.022  &  0.033&0.015 & -0.039&0.029 & -0.029&0.025 & -0.020&0.021 &  99.9  &  99.8  &  99.9  &  99.8 \\
 \ngc4833                  & -0.056&0.057  &  0.123&0.037 &  0.038&0.040 &  0.022&0.051 &  0.017&0.056 &  96.0  &  95.8  &  91.6  &  96.6 \\
 \ngc5272                  & -0.081&0.048  & -0.044&0.019 & -0.065&0.042 & -0.053&0.015 & -0.047&0.019 &  95.9  &  99.6  &  95.5  &  99.5 \\
 \ngc5904                  & -0.109&0.031  &  0.135&0.043 & -0.023&0.026 & -0.002&0.035 & -0.003&0.031 &  90.5  & 100.0  &  88.1  & 100.0 \\ 
 \ngc5927                  & -0.114&0.054  &  0.020&0.024 & -0.037&0.075 & -0.123&0.041 & -0.121&0.020 & 100.0  &  99.8  & 100.0  &  99.8 \\
 \ngc5986                  & -0.022&0.016  &  0.113&0.037 & -0.010&0.025 & -0.001&0.023 & -0.009&0.026 &  99.5  & 100.0  &  99.2  &  99.9 \\
 \ngc6093                  & -0.162&0.034  &  0.012&0.022 & -0.080&0.025 & -0.029&0.027 & -0.053&0.022 &  99.8  &  92.2  &  99.9  &  90.6 \\
 \ngc6171                  & -0.095&0.049  &  0.028&0.053 & -0.122&0.054 & -0.096&0.048 & -0.086&0.045 &  92.9  &  99.5  &  94.7  &  99.7 \\
 \ngc6205                  & -0.130&0.037  &  0.006&0.030 & -0.081&0.024 & -0.078&0.021 & -0.082&0.019 &  97.9  &  99.9  &  99.0  &  99.9 \\
 \ngc6304                  & -0.160&0.061  &  0.009&0.056 & -0.069&0.024 & -0.027&0.022 & -0.010&0.026 &  97.3  &  99.0  &  97.0  &  99.6 \\
 \ngc6352                  & -0.168&0.042  & -0.025&0.030 & -0.131&0.025 & -0.121&0.031 & -0.108&0.032 &  99.5  &  99.1  &  99.6  &  98.9 \\
 \ngc6362                  & -0.042&0.041  &  0.108&0.034 &  0.027&0.042 &  0.029&0.042 &  0.049&0.042 & 100.0  &  94.3  & 100.0  &  96.2 \\
 \ngc6388                  &  0.017&0.052  &  0.016&0.019 & -0.094&0.045 & -0.154&0.062 & -0.066&0.071 &  91.1  & 100.0  &  94.5  & 100.0 \\
 \ngc6441                  & -0.235&0.034  & -0.011&0.037 & -0.040&0.025 & -0.038&0.023 & -0.014&0.032 & 100.0  & 100.0  & 100.0  & 100.0 \\
 \ngc6624                  & -0.207&0.041  &  0.023&0.027 & -0.063&0.049 & -0.068&0.025 & -0.058&0.022 &  98.7  & 100.0  &  98.6  & 100.0 \\
 \ngc6637                  & -0.092&0.022  &  0.111&0.054 & -0.044&0.074 & -0.022&0.037 &  0.001&0.037 &  83.0  & 100.0  &  89.8  & 100.0 \\
 \ngc6652                  & -0.144&0.025  &  0.031&0.025 & -0.064&0.046 & -0.056&0.036 & -0.051&0.031 &  99.2  &  99.6  &  99.2  &  99.8 \\
 \ngc6681                  & -0.046&0.035  &  0.076&0.027 & -0.016&0.036 & -0.002&0.038 & -0.007&0.043 &  99.1  &  93.6  &  99.2  &  93.6 \\
 \ngc6715                  & -0.037&0.111  &  0.051&0.066 & -0.023&0.060 & -0.048&0.027 & -0.042&0.034 &  99.1  &  99.2  &  98.9  &  99.5 \\
 \ngc6723                  &  0.008&0.024  &  0.121&0.018 &  0.005&0.016 & -0.021&0.025 & -0.004&0.022 &  98.2  &  98.7  &  97.8  &  98.6 \\
 \ngc6752\tnote{*}         & -0.112&0.016  &  0.030&0.019 & -0.043&0.021 & -0.066&0.022 & -0.072&0.022 &  99.5  &  99.9  &  99.9  & 100.0 \\
 \ngc6934                  & -0.134&0.028  &  0.044&0.024 & -0.072&0.035 & -0.055&0.038 & -0.054&0.036 &  97.4  &  99.8  &  98.1  &  99.8 \\
 \ngc6981                  & -0.108&0.026  &  0.063&0.035 & -0.037&0.037 & -0.038&0.032 & -0.030&0.038 &  97.2  &  99.8  &  95.5  &  99.5 \\
 \ngc7078                  & -0.081&0.016  &  0.067&0.068 & -0.100&0.021 & -0.086&0.025 & -0.095&0.020 &  87.8  &  98.5  &  90.7  &  98.8 \\
\hline
\end{tabular}
\begin{tablenotes}
\item[*]{The GCs \ngc104 and \ngc6752 have not been observed in the F438W filter but in the similar passband filter F435W \citepalias[see][]{Piotto15}.}
\end{tablenotes}
\end{threeparttable}
\end{table*}

\begin{figure*}
\includegraphics[width=\textwidth]{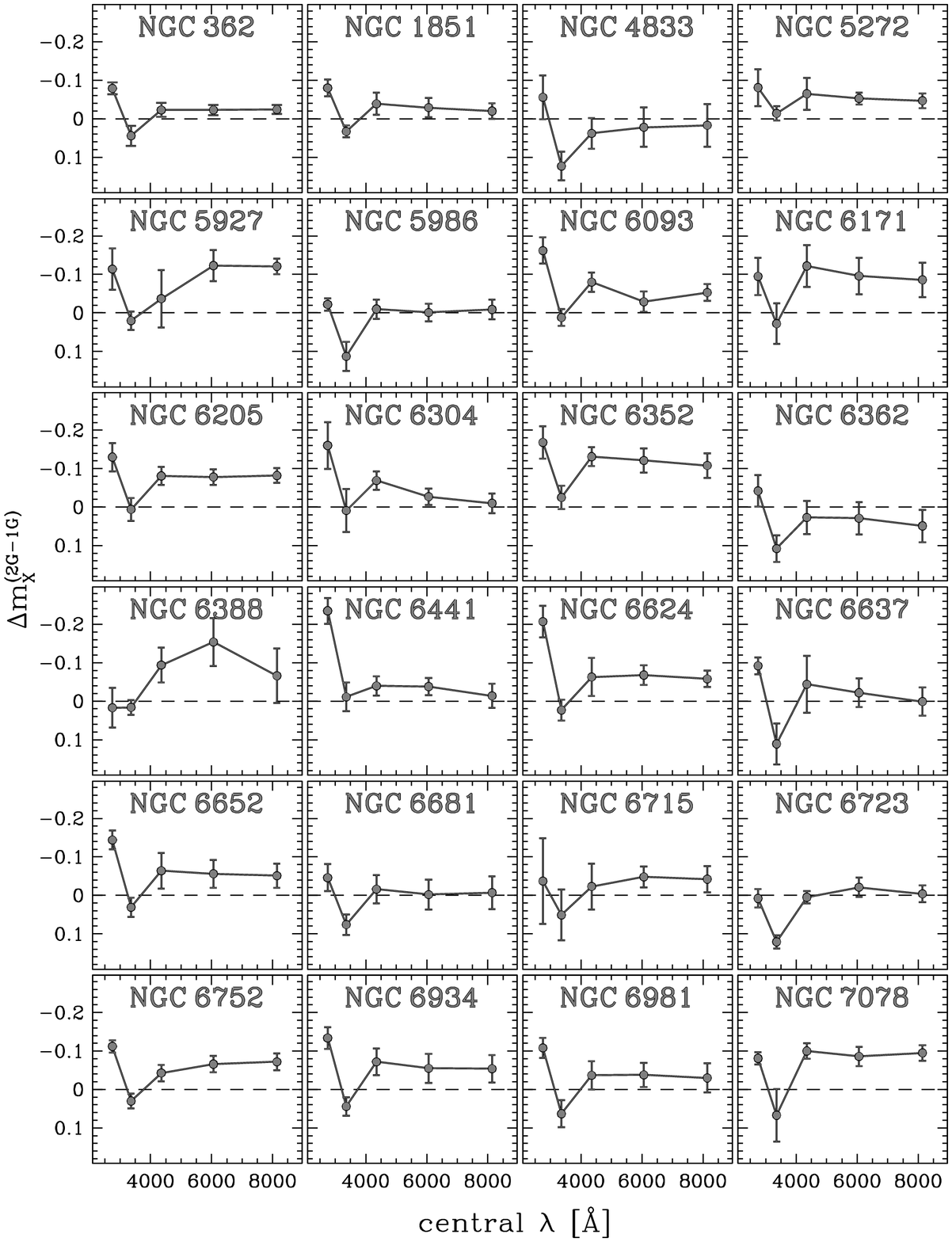}
\caption{As in Figure~\ref{fig:delta}, for all the clusters other than \ngc104,
with a significance value $\geq 90\%$ in the F814W band.\label{fig:dbumps}}
\end{figure*}

Figure~\ref{fig:dbumps} is similar to Figure~\ref{fig:delta} and shows the
magnitude difference between the RGBB of the 2G and 1G stars of all the GCs in
Table~\ref{tab:tab1} (except \ngc104) as a function of the central wavelength of
the filter X. 

A look at the plot reveals that all the GCs show comparable trends in the
$\rm \Delta m_{X}^{(2G,1G)}$ vs.\,central $\lambda$ diagram. In all bands but F336W
the 2G RGBB is brighter than the 1G RGBB, in close analogy with what we observe
for \ngc104. \ngc4833 and \ngc6362 are possible exceptions because the 2G RGBB
is fainter than the 1G RGBB in the three optical bands. The F438W, F606W, and
F814W magnitude differences between the two main RGBBs of all the clusters are
very similar to each other and are smaller than $\sim 0.1$ mag.

\subsection{The effects of C, N, and O on the RGB Bump luminosity}\label{subsec:CNO}
\begin{figure*}
\includegraphics[width=0.87\textwidth]{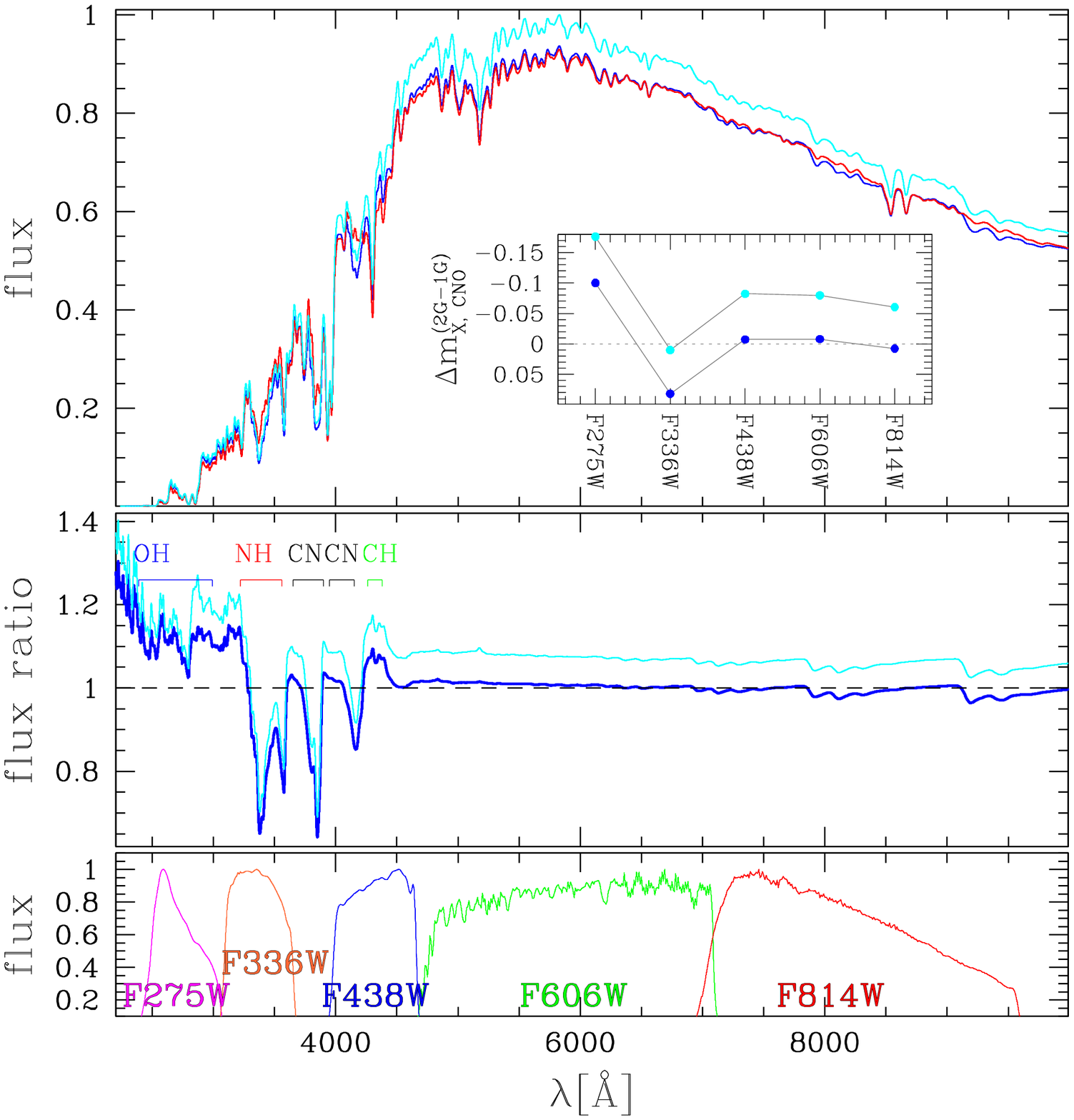}
\caption{Comparison of 1G and 2G synthetic spectra. The red spectrum
plotted in the upper panel has the same chemical composition and stellar
parameters as a 1G RGB Bump star with $\feh=-0.75$ and age$=12.5$ Gyr. The blue
and the cyan spectra have the same metallicity and age as the red spectrum but
the C, N, and O content of 2G stars. The cyan spectrum corresponds to a RGB Bump
star enhanced in helium by $\rm \Delta Y = 0.074$ with respect
to the other two spectra, which have $\rm Y=0.256$. The ratio
of the fluxes of the cyan and blue synthetic spectrum with respect to the red
one are plotted in the middle panel as a function of the wavelength, while the
transmission curves of the ACS/WFC and WFC3/UVIS filters used in this paper are
shown in the lower panel. The inset in the upper panel shows the magnitude
difference between the blue and cyan 2G spectra and the 1G synthetic
spectrum in the five \hst\ bands used in this paper. See text for
details.\label{fig:spettri}} 
\end{figure*}

To investigate the physical reasons responsible for the observed magnitude
difference between the RGBB of 2G and 1G stars we compared the $\rm \Delta
m_{X}^{(2G,1G)}$ values with the magnitude of RGBB stars derived from synthetic
spectra with appropriate chemical composition. To do this, we extended to the
RGBB the procedure used in our previous papers to characterize the multiple
populations along the MS and RGB. The main steps of our analysis are illustrated
in Figure~\ref{fig:spettri} for an RGBB star with $\feh=-0.75$ (appropriate for
\ngc104).

First, we determined the \teff\ and gravity ($\log{g}$) corresponding to the
RGBB by using isochrones from the Bag of Stellar Tracks and Isochrones (BaSTI)
database~\footnote{\url{http://basti.oa-teramo.inaf.it}}
\citep{Pietrinferni04,Pietrinferni06,Pietrinferni09} with $\feh=-0.75$ and age
of 12.5 Gyr. 
 
Then, we simulated two synthetic spectra with these atmospheric parameters but
different C, N, O abundances that resemble the chemical composition of 1G and 2G
stars inferred from high-resolution spectroscopy. Specifically, we assumed for
the 1G spectrum [C/Fe]$=0.0$, [N/Fe]$=0.1$, and [O/Fe]$=0.3$, while for the
comparison spectrum we used [C/Fe]$=-0.3$, [N/Fe]$=0.9$, [O/Fe]$=0.0$. These
values are close to the average C, N abundances of the 1G and 2G RGB stars in
\ngc104 obtained by \citet{Marino16} and to the average O abundance derived by
\citet{Carretta09} for bright RGB stars. We assumed that these spectra have
primordial helium content (Y=0.256).

To investigate the effect of helium variation on the luminosity of the RGBB, we
simulated a third spectrum with the same C, N, and O abundance as 2G stars but
with an higher helium content $\rm (Y=0.33)$. The corresponding
atmospheric parameters are derived from BaSTI isochrones.

The ATLAS12 and SYNTHE codes \citep{Castelli04,Kurucz05,Sbordone07} are used to
generate the synthetic spectra in the wavelength interval between 2,000 and
10,000 \AA. In the upper panel of Figure~\ref{fig:spettri} is shown the
comparison between the spectra of the 1G (red) and the 2G (blue) star with
primordial helium content and the 2G star with $\rm Y=0.33$ (cyan). The
corresponding flux ratios are plotted in the middle panel as a function of
$\lambda$. For the sake of completeness we show in the bottom panel the
normalized transmission curves of the \hst\ filters used in this work.

The flux of each spectrum has been convolved with the transmission curve of the
WFC3/UVIS F275W, F336W and F438W filters and with the ACS/WFC F606W and F814W
filters, to derive the corresponding synthetic magnitudes and the $\rm \Delta
m_{X,CNO}^{(2G,1G)}$ values. The results are shown in the inset of the upper
panel of Figure~\ref{fig:spettri} where we plot the derived values of $\rm
\Delta m_{X,CNO}^{(2G,1G)}$ for X = F275W, F336W, F438W, F606W and F814W.

From the comparison of the 2G and 1G spectra with primordial helium, we found
that C, N, and O variations strongly affect the F275W and F336W magnitudes
mostly through the OH and NH molecular bands and result in large magnitude
differences between the 2G and the 1G spectrum. Carbon and nitrogen are also
responsible for significant flux variation in the wavelength region covered by
the F438W filter due to the absorption of CN bands. The effects of C, N, and O
are less pronounced in the optical bands where the magnitude difference between
the two spectra is $\lesssim 0.01$.

In contrast, the comparison between the helium-rich and primordial
helium $\rm (Y=0.256)$ 2G star spectra shows that helium
enhancement mostly affects the bolometric luminosity of the star, resulting in a
magnitude difference of $\sim 0.07$ mag in the F606W and F814W
bands.

The fact that optical magnitudes are poorly affected by C, N, O
variations but are sensitive to helium, demonstrates that the magnitude
difference between 2G and 1G stars in optical bands provide a strong constraint
of their relative helium content.

\begin{figure*}
\includegraphics[width=0.85\textwidth]{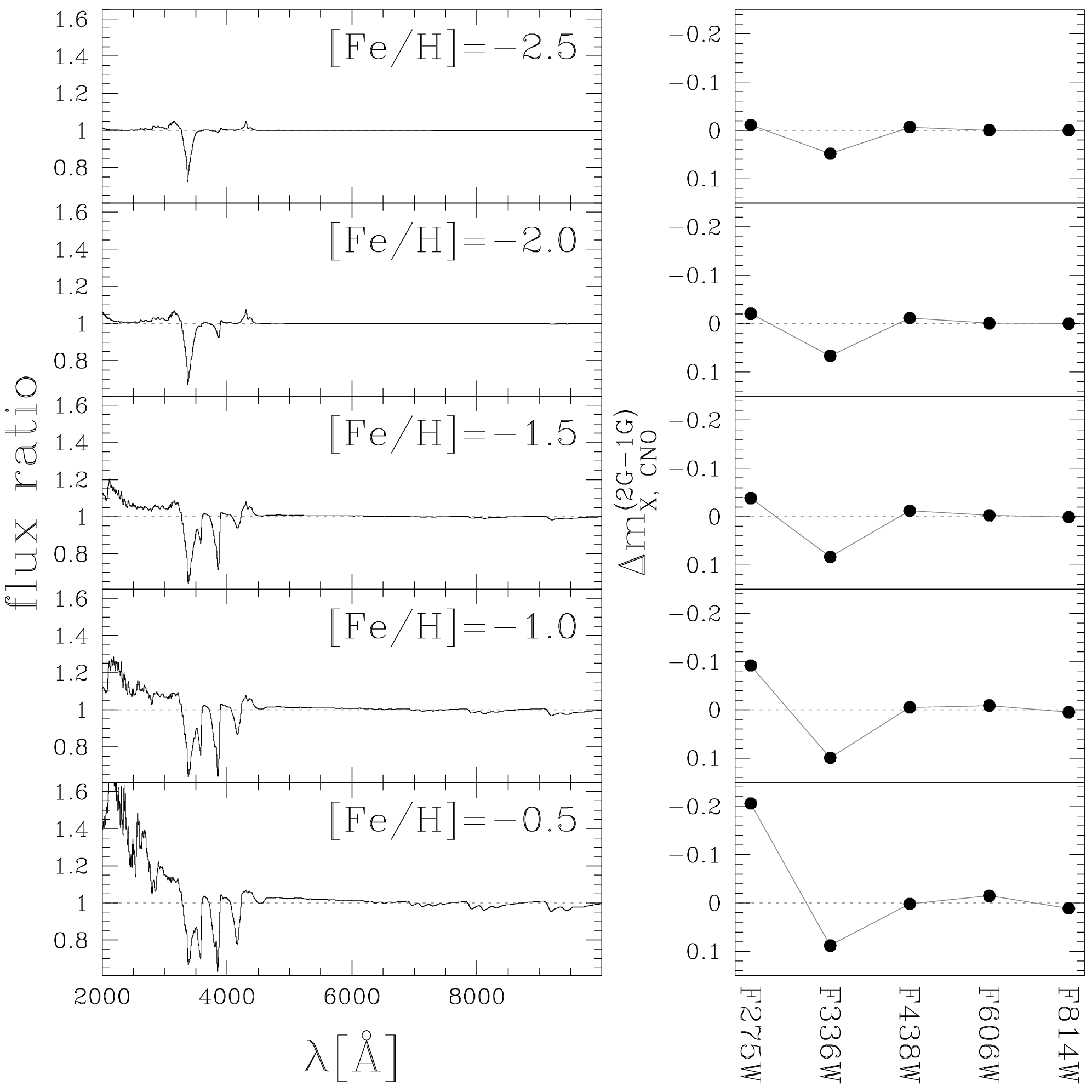}
\caption{Flux ratio of 2G to 1G reference spectra for RGBB stars with different
metallicity (left panels). The corresponding magnitude difference is plotted in
the right panels for the five filters used in this paper.\label{fig:DFLUX}}
\end{figure*}

To investigate the effect of light-element variations in the spectra of
clusters with different metallicity, in Figure~\ref{fig:DFLUX} we extended the
analysis to the RGB stars of five 12.5 Gyr-old GCs with $\feh=-2.5$, $-2.0$,
$-1.5$, $-1.0$, $-0.5$. In all the cases we assumed for the two synthetic
spectra the C, N, and O proportions used to derive the spectra plotted in
Figure~\ref{fig:spettri}. The flux ratios of the spectra with the chemical
composition of 2G stars to those with the chemical composition of 1G stars are
plotted in the left panels while the right panels show the corresponding $\rm
\Delta m_{X,CNO}^{(2G,1G)}$ values obtained for the five \hst\ filters used in
this paper.

In the left panels we observe that the strength of the molecular absorption
bands increases with metallicity. This trend is reflected, in the right panels,
in the variation of the $\rm \Delta m_{X,CNO}^{(2G,1G)}$, mostly visible in the
F275W and F336W bands, as discussed above. Indeed, the F275W magnitude
difference between the two spectra, that has a value of about $-0.01$ mag at
$\feh=-2.5$, becomes smaller than $-0.2$ mag at $\feh=-0.5$, while the F336W
magnitude difference steadily increases from $\sim 0.04$ mag at $\feh=-2.5$ up to
$\sim 0.1$ mag for $\feh \sim -1.0$ and then flattens to $\sim 0.01$ towards
$\feh=-0.5$.

In contrast, the magnitude differences due to C, N, and O variations are
typically small in the F438W, F606W and F814W filters. In particular, $\rm
\Delta m_{F606W,CNO}^{(2G,1G)}$ and $\rm \Delta m_{F814W,CNO}^{(2G,1G)}$ are
consistent with zero in metal-poor RGBB stars and their absolute values are
significantly smaller than 0.01 mag for $\feh\lesssim -0.8$. Despite being
significantly affected by the strong CH bands, the F438W magnitude difference
between the two spectra is quite small and is slightly larger than 0.01 mag only
for $-2.2 \lesssim\feh\lesssim -1.2$. 

A visual inspection at Figs.~\ref{fig:delta} and \ref{fig:dbumps} reveals that
several clusters, including metal-poor GCs, exhibit large $\rm \Delta
m_{X}^{(2G,1G)}$ values in contrast with what we expect from light-element
differences alone between 2G and 1G stars.

\begin{figure}
\includegraphics[width=\columnwidth]{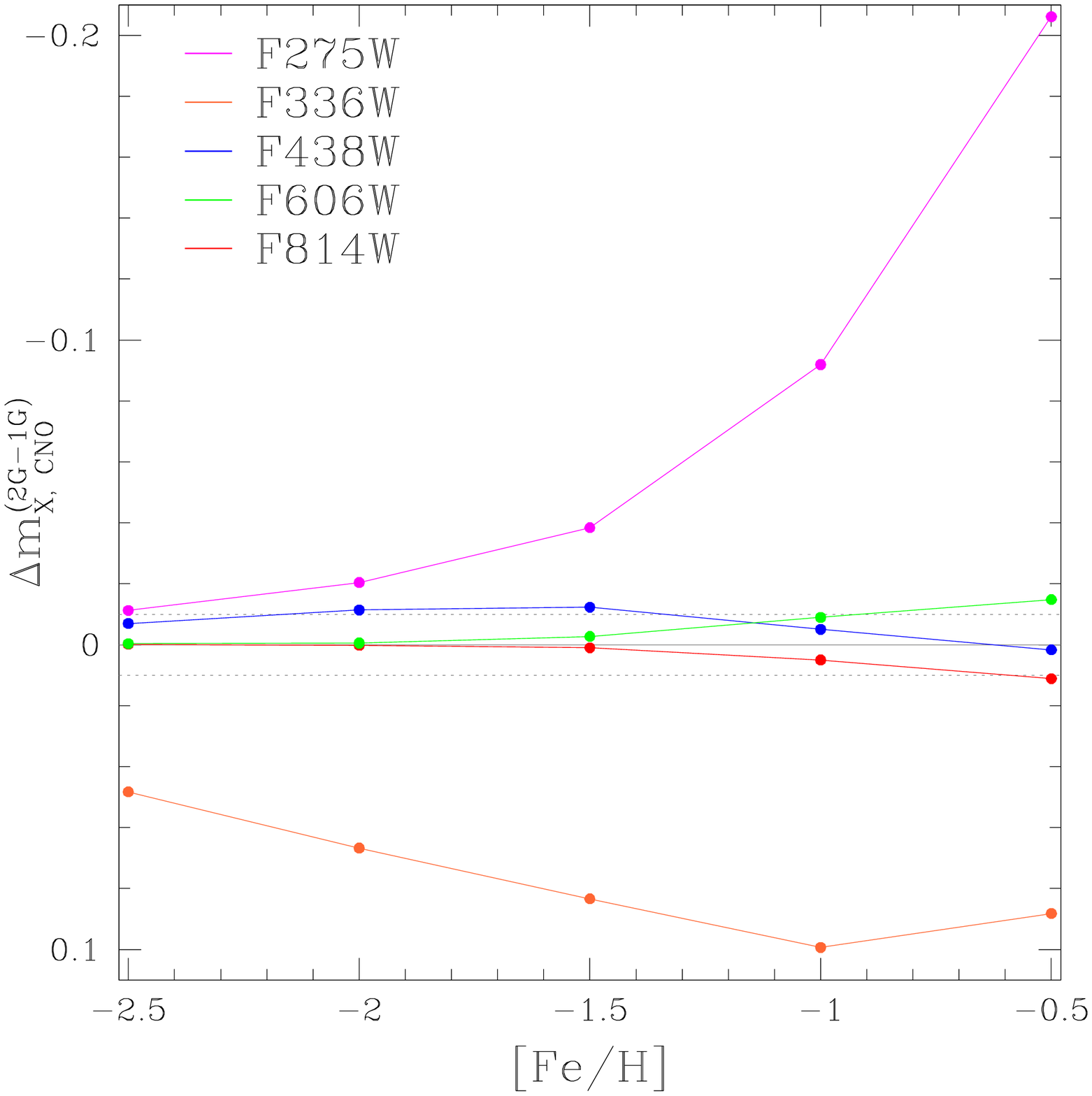}
\caption{Magnitude difference between the comparison and the reference spectra
of RGBB stars as a function of the metallicity. The five filters analysed in
this paper are represented with different colours as quoted in the figure. The
horizontal dotted lines are located at
$\rm \Delta m_{X,CNO}^{(2G,1G)} = +/-\,0.01$.\label{fig:DMAG}}
\end{figure}

\subsection{The relative helium abundance of 1G and 2G stars}\label{subsec:relY}
The analysis of the synthetic spectra demonstrates that C, N, and O variations
alone are not able to reproduce the long-wavelength (F438W, F606W and F814W)
differences between 1G and 2G stars. As a consequence, the observed magnitude of
the RGBBs must also depend on the helium abundance of 1G and 2G stars and can be
expressed as: 
\begin{equation} \label{eq:eq1} 
\rm \Delta m_{X}^{(2G,1G)} = \Delta m_{X,CNO}^{(2G,1G)} + \Delta m_{X,He}^{(2G,1G)}
\end{equation} 
where the last two terms of this relation indicate the contribute of the C, N, O
and helium variations, respectively. In this section, we infer the helium
difference between 2G and 1G stars in each cluster by comparing the observed
RGBB magnitude separation with the quantities $\rm \Delta m_{X,CNO}^{(2G,1G)}$
and $\rm \Delta m_{X,He}^{(2G,1G)}$ predicted by theoretical models with
appropriate chemical composition.

The quantity $\rm \Delta m_{X,CNO}^{(2G,1G)}$ has been derived for each cluster
by using the procedure described in Section~\ref{subsec:CNO}. As the relative C
and N abundance of 1G and 2G stars is not available for most of the analysed
GCs, modelling their effect on the RGBB luminosity is actually one of the main
challenges of our analysis. For simplicity, we assumed for all the GCs the
abundance of C, N, and O inferred from high-resolution spectroscopy of stars in
\ngc104 by \citet{Carretta09} and \citet{Marino16}. Because of this
approximation, to minimize the uncertainty on the helium determination, we
limited our analysis to those filters and clusters where the contribution of the
C, N, and O on the RGBB luminosity is negligible, namely F606W and
F814W. Indeed, in these two filters, C, N, and O variations produce RGBB
luminosity differences of less than 0.01 mag for metallicities lower than $\feh
\sim -1.0$. Therefore, the following analysis has been performed only in the
F606W and F814W bands and for the GCs with $\feh \lesssim -1.0$. \ngc104, for
which we have accurate C, N, O abundances from spectroscopy, is the only
metal-rich cluster included in the analysis.

The F438W magnitude seems also poorly affected by the adopted C, N, and
O variations. However, due to the presence of strong CN and CH bands, we prefer
to not infer the helium abundance from this filter.

To estimate the quantity $\rm \Delta m_{X,He}^{(2G,1G)}$ that best matches the
observations, and derive the relative helium abundance between 2G and 1G stars,
we used the models of the BaSTI database.

For each cluster we calculated a grid of alpha-enhanced isochrones
(\afe$\ =0.4$): a reference isochrone with standard helium content,
$\rm Y \approx 0.25$, and a set of helium-enhanced isochrones that, for helium
abundances not available in the original grid, have been computed for
interpolation among the available grid points. Metallicity and age values were
taken respectively from \citet[][2010 ed.]{Harris96} and \citet{Dotter10}. 

To determine the helium difference between 2G and 1G stars from F814W stellar
magnitudes we first generated a grid of synthetic CMDs for the RGB stars.
Specifically, we simulated a CMD corresponding to the reference isochrone and
100 CMDs derived from isochrones enhanced in helium by $\rm \Delta Y$ with
respect to the standard value. For each helium-enhanced isochrone, $i$, we
assumed ${\rm \Delta Y}_i$ ranging from 0.000 to 0.100 in steps of 0.001.

Each synthetic CMD was derived by using 200,000 artificial stars
\citep{Anderson08}, to account for the observational errors in colour and
magnitude. Moreover, we assumed for the LF of the reference and of each
helium-rich synthetic CMD, $i$, the same slope of the  corresponding observed
LF.

We determined the RGBB magnitude for each couple of synthetic CMDs, by using the
same procedure described in Section~\ref{sec:RGBLF} for real stars. Then, we
estimated the corresponding magnitude difference, ${\rm \Delta m}_{i\ \rm
F814W,He}^{\rm (2G,1G)}$, and assumed as the best estimate of the helium difference between
2G and 1G stars the value of ${\rm \Delta Y}_i$ that provides: 
${\rm \Delta m}_{i\ \rm F814W,He}^{\rm (2G,1G)} = \rm \Delta m_{F814W}^{(2G,1G)} -
\Delta m_{X,CNO}^{(2G,1G)}$. We applied the same method to infer the helium
abundance from the F606W band.

\begin{figure*}
\includegraphics[width=0.85\textwidth]{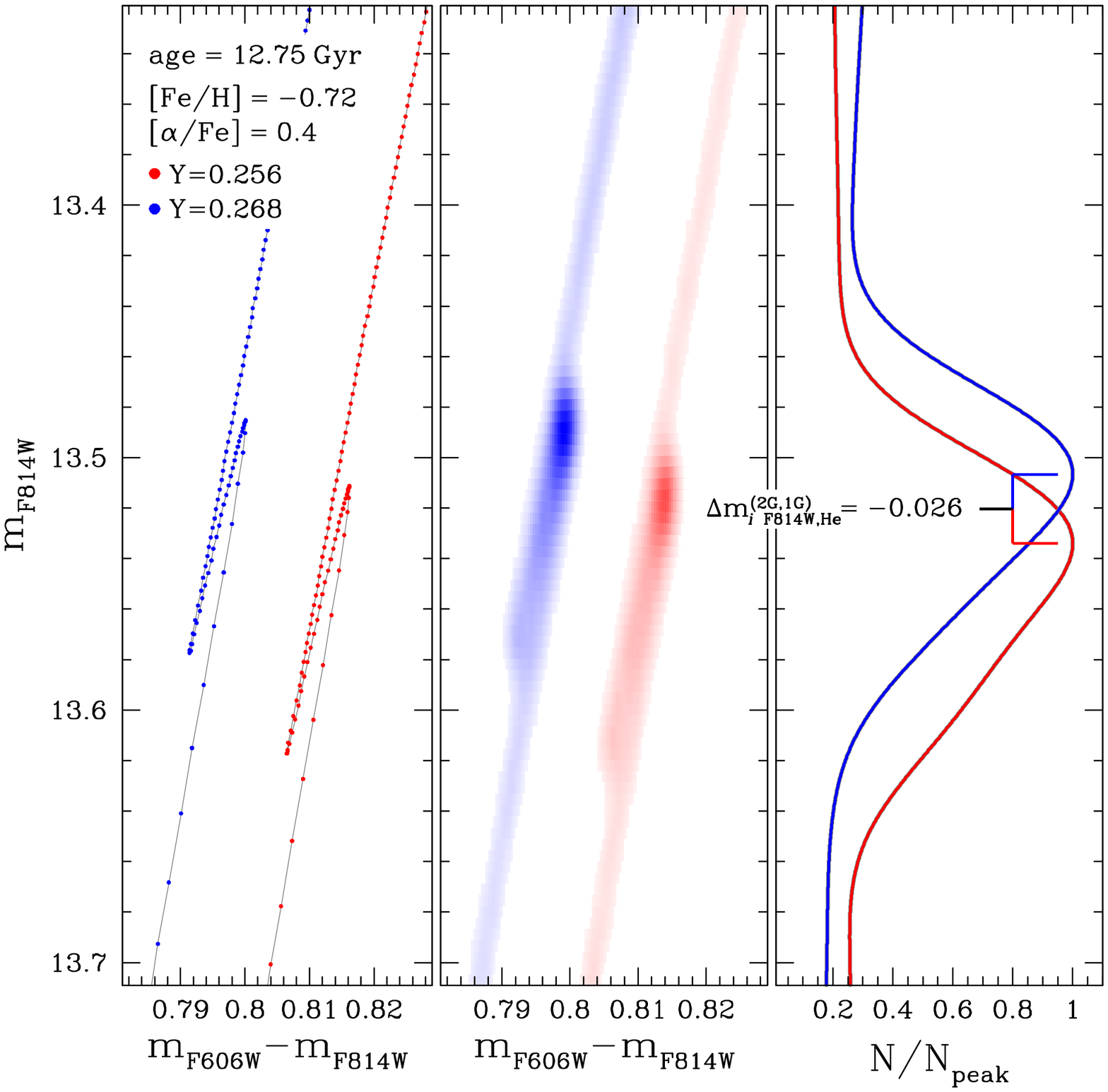}
\caption{Procedure to estimate the helium abundance difference, $\rm \Delta Y$,
between the 1G and 2G stars of \ngc104 in the F814W band. {\it Left panel}:
$\rm m_{F814W}$ vs.\,$\rm m_{F606W}-m_{F814W}$ CMD of an alpha-enhanced
BaSTI isochrone with He-standard ($\rm Y=0.256$, red points) and He-enhanced
($\rm Y=0.268$, blue points) content. The latter coincides with our best estimate of
the helium content of the 2G stars of \ngc104 in the F814W band. Since the two
models almost attain the same colour, the He-enhanced isochrone has been shifted
of $-0.15$ mag along the colour axis. The adopted values of age, \feh, and \afe\
are also quoted in the figure.  {\it Central panel}: $\rm m_{F814W}$
vs.\,$\rm m_{F606W}-m_{F814W}$ Hess diagram of the synthetic RGBs obtained
from the $\rm Y=0.256$ and $\rm Y=0.268$ models (see text for details). Along each
sequence, represented with the same colour code of the corresponding isochrone,
is visible an overdensity that represents the RGB Bump. {\it Right panel}:
Kernel density estimate of the synthetic LFs relative to the He-standard (red
curve) and He-enhanced (blue curve) model. Each curve has been normalized to the
corresponding peak value. The magnitude difference between the two peaks, ${\rm
\Delta}_{i\ \rm F814W,He}^{\rm (2G,1G)}=-0.026$, has also been
reported.\label{fig:synt}} 
\end{figure*}

For example, the procedure for the estimate of the $\rm \Delta Y$ between the 2G
and 1G stars of \ngc104 in the F814W band is displayed in Figure~\ref{fig:synt}.
The left panel shows the $\rm m_{F814W}$ vs.\,$\rm m_{F606W}-m_{F814W}$ CMD
centred on the approximate location of the RGBB of the two adopted models: the
red points indicate the reference isochrone, with $\rm Y=0.256$ while the blue ones
the He-enhanced isochrone. The latter has a helium content of $\rm Y=0.268$,
corresponding to our best estimate of the helium abundance of the 2G stars of
\ngc104 in the F814W band, and has been obtained by linearly interpolating the
BaSTI models with $\rm Y=0.256$ and $\rm Y=0.300$. Both the isochrones have an age of
12.75 Gyr and a metallicity of $\feh=-0.72$. Since the two models almost attain
the same colour, for the sake of clarity we added $-0.15$ mag to the colour of the
He-enhanced isochrone.

The central panel displays the Hess diagram of the synthetic RGBs relative to
the $\rm Y=0.256$ and $\rm Y=0.268$ models. The RGBB is indicated by the overdensity
visible in each sequence, plotted with the same colour code of the corresponding
isochrone. 

Both the LFs have been normalized to the peak value of the corresponding kernel
density estimate, plotted as a solid curve with the same colour code of the
relative model, and the magnitude difference between the peaks of the two
curves, ${\rm \Delta m}_{i\ \rm F814W,He}$ has also been reported. In
the case of \ngc104, the magnitude difference due to the C, N, and O variations
is $\rm \Delta m_{X,CNO}^{(2G,1G)}=0.008$ and the observed magnitude
difference is $\rm \Delta m_{F814W}^{(2G,1G)}=-0.018$. Hence the adopted value
of $\rm Y=0.268$, which corresponds to ${\rm \Delta m}_{i\ \rm
F814W,He}^{\rm (2G,1G)}=-0.026$, satisfies equation~\ref{eq:eq1} and represents the
best estimate for the $\rm \Delta Y_{F814W}$.

In Table~\ref{tab:tab2} we provide, for the selected 18 clusters, the adopted
metallicity and age values, the $\rm \Delta Y$ estimates in the F606W and F814W
bands and their average value, with the corresponding standard error, in columns
2, 3, 4, 5, 6, respectively.

We notice that for each cluster the $\rm \Delta Y_{F606W}$ and the $\rm \Delta
Y_{F814W}$ are consistent at $1\sigma$ level, with an average difference of
$0.001\pm0.003$. Therefore, we considered the weighted mean of the two values,
$\rm \langle \Delta Y \rangle$, as our best estimate for the helium difference
between the 2G and 1G stars in each cluster.

\begin{table*}
\centering
\caption{List of the 18 selected GCs for which we estimated the $\rm \Delta Y$ by
using BaSTI theoretical models. Column 6 gives the adopted estimate,
$\rm \langle \Delta Y \rangle$, obtained as the weighted mean of the $\rm \Delta
Y_{F606W}$ (column 4) and $\rm \Delta Y_{F814W}$ (column 5)
values.\label{tab:tab2}}
\normalsize
\begin{tabular}{lccr@{$\,\pm\,$}lr@{$\,\pm\,$}lr@{$\,\pm\,$}l}
\hline
Cluster & \feh & age (Gyr) & \multicolumn{2}{c}{$\rm \Delta Y_{F606W}$} & \multicolumn{2}{c}{$\rm \Delta Y_{F814W}$} & \multicolumn{2}{c}{$\rm \langle \Delta Y \rangle$} \\
\hline
  \ngc104  & -0.72 & 12.75 &  0.009&0.005 &  0.012&0.007 &  0.010&0.004 \\         
  \ngc362  & -1.26 & 11.50 &  0.006&0.004 &  0.008&0.004 &  0.007&0.003 \\        
  \ngc1851 & -1.18 & 11.00 &  0.008&0.008 &  0.007&0.006 &  0.007&0.005 \\
  \ngc4833 & -1.85 & 13.00 & -0.006&0.016 & -0.005&0.018 & -0.006&0.012 \\        
  \ngc5272 & -1.50 & 12.50 &  0.013&0.004 &  0.012&0.005 &  0.013&0.003 \\        
  \ngc5904 & -1.29 & 12.25 & -0.002&0.012 &  0.002&0.010 &  0.000&0.008 \\
  \ngc5986 & -1.59 & 13.25 &  0.000&0.006 &  0.002&0.006 &  0.001&0.004 \\       
  \ngc6093 & -1.75 & 13.50 &  0.007&0.008 &  0.014&0.006 &  0.011&0.005 \\        
  \ngc6171 & -1.02 & 12.75 &  0.029&0.016 &  0.031&0.015 &  0.030&0.011 \\ 
  \ngc6205 & -1.53 & 13.00 &  0.020&0.006 &  0.022&0.005 &  0.021&0.004 \\        
  \ngc6362 & -0.99 & 12.50 & -0.014&0.014 & -0.015&0.014 & -0.014&0.010 \\        
  \ngc6681 & -1.62 & 13.00 &  0.000&0.009 &  0.002&0.010 &  0.001&0.007 \\        
  \ngc6715 & -1.49 & 13.25 &  0.013&0.007 &  0.012&0.009 &  0.013&0.006 \\
  \ngc6723 & -1.10 & 12.75 &  0.005&0.008 &  0.003&0.007 &  0.004&0.005 \\      	  
  \ngc6752 & -1.54 & 12.50 &  0.017&0.006 &  0.019&0.006 &  0.018&0.004 \\        
  \ngc6934 & -1.47 & 12.00 &  0.014&0.011 &  0.016&0.010 &  0.015&0.007 \\ 
  \ngc6981 & -1.42 & 12.75 &  0.012&0.010 &  0.009&0.011 &  0.011&0.007 \\        
  \ngc7078 & -2.37 & 13.25 &  0.030&0.010 &  0.033&0.008 &  0.032&0.006 \\        
\hline
\end{tabular}
\end{table*}

The mean $\rm \Delta Y$ values indicate that, in all the analysed clusters, the
2G stars are helium enhanced with respect to the 1G stars by less than $\sim
0.035$ in mass fraction. Moreover in some clusters, namely \ngc4833, \ngc5986,
\ngc6681, and \ngc6723, the 1G and 2G stars have the same helium abundance at
$\sim 1\sigma$ level. 

It is worth to notice that, for each cluster, the $\rm \langle \Delta Y
\rangle$ value was derived by assuming that the 1G and 2G stars are coeval, as
pointed out by \citet{Marino12} for the GC \ngc6656 and by
\citetalias{Nardiello15} for the GC \ngc6352. The typical error affecting the
estimate of the relative age between the 1G and 2G is between 100 and 300 Myr.
For this reason, we decided to relax the condition of coeval stellar generation
and repeated the computation of the $\rm \Delta Y$ by assuming a population of
2G stars 100 Myr younger than 1G stars. This assumption has no
significant impact on our estimates of $\rm \langle \Delta Y \rangle$. For
example, we obtain, for the cluster \ngc104, a difference in $\rm \langle \Delta
Y \rangle \lesssim 0.001$, which is negligible for our purposes. Moreover, we
verified that the uncertainty on the age of GCs also does not affect
significantly the $\rm \langle \Delta Y \rangle$ estimates. Again, in the case
of \ngc104, an age difference of $\pm 0.75$ Gyr, which is the typical error of
the age from \citet{Dotter10}, corresponds to a difference in $\rm \langle
\Delta Y \rangle \lesssim 0.001$.
  
\begin{figure}
\includegraphics[width=\columnwidth]{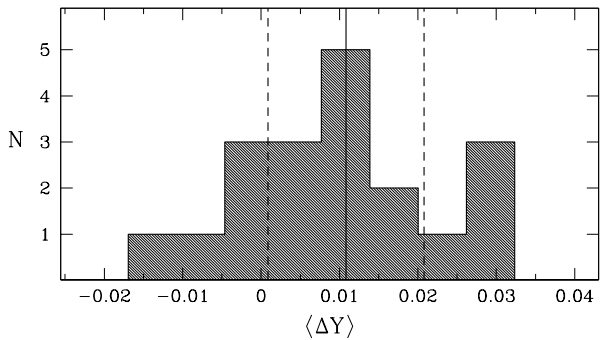}
\caption{Distribution of the $\rm \langle \Delta Y \rangle$ values of the
clusters listed in column 6 of Table~\ref{tab:tab2} plus \ngc2808. The histogram
was obtained by using bins of $\rm \langle \Delta Y \rangle = 0.006$. The
vertical solid line at $\rm \langle \Delta Y \rangle = 0.011$ and the two dashed
lines at $\rm \langle \Delta Y \rangle = 0.001$ and $\rm \langle \Delta Y
\rangle = 0.021$ mark, respectively, the mean and the $+/-\,1\sigma$ of the
distribution.\label{fig:DY_dist}}
\end{figure}

Figure~\ref{fig:DY_dist} displays the histogram distribution of the $\rm
\langle \Delta Y \rangle$ values for the selected GCs and \ngc2808 from
\citetalias{Milone15b}~\footnote{\citetalias{Milone15b} derived the relative
helium abundance for the four main populations, namely B--E, of \ngc2808 by
following the same method used in this paper. We estimated the value of $\rm
\langle \Delta Y \rangle=0.032\pm0.008$ for \ngc2808 by assuming that population
B corresponds to the 1G while the 2G is composed of the populations C, D, and E
(see \citetalias{Milone15b} and \citetalias{Milone17} for details).}. The
histogram was built by using bins of $\rm \Delta Y=0.006$. The distribution
clearly shows that, for the clusters in our sample, 2G stars are helium enriched
with respect to the 1G stars with a mean value $\rm \langle \Delta Y \rangle=
0.011\pm0.002$, marked by the vertical solid line. The two dashed lines at $\rm
\langle \Delta Y \rangle=0.001$ and $\rm \langle \Delta Y \rangle=0.021$
indicate the points at $+/-\,1\sigma$, respectively, with $\sigma=0.010$.

\begin{figure*}
\includegraphics[width=\textwidth]{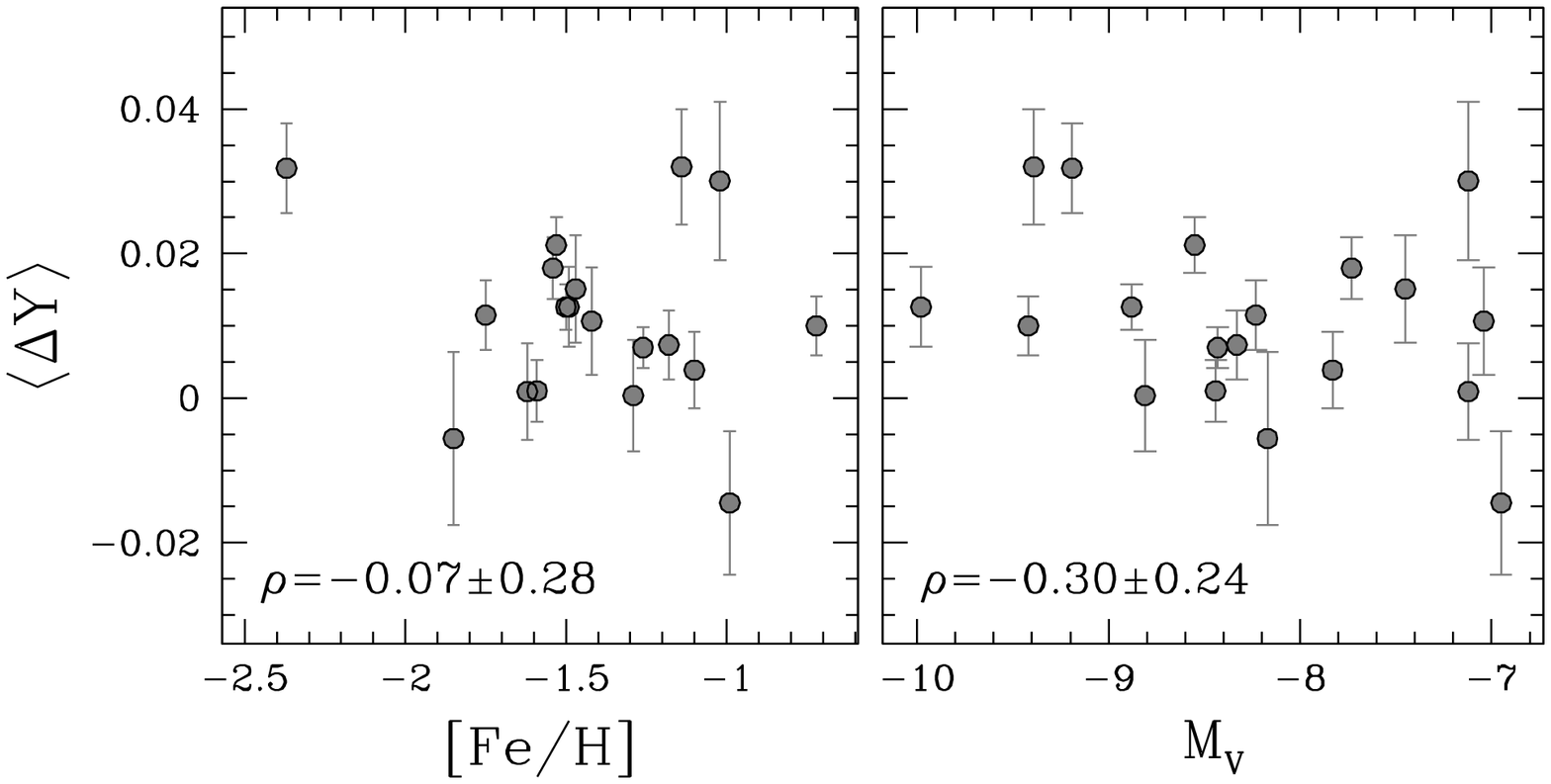}
\caption{Plot of $\rm \langle \Delta Y \rangle$ vs.\,\feh\ (left panel) and
vs.\,$\rm M_V$ (right panel) of the clusters listed in Table~\ref{tab:tab2} plus
\ngc2808. The values of the Spearman's rank correlation coefficient ($\rho$) are
indicated in each plot.\label{fig:corr_2}}
\end{figure*}

In \citetalias{Milone17} we show that the RGB width in the $\rm m_{F275W}-m_{F814W}$
colour and in the $\rm C_{F275W,F336W,F438W}$ pseudo-colour correlate with the
cluster absolute luminosity and with the metallicity. Similarly, the fraction of
2G stars with respect to the total number of stars correlates with the cluster
absolute luminosity thus indicating that the incidence and complexity of the
multiple-population phenomenon both increase with cluster mass.

Similarly to what we have done in \citetalias{Milone17}, we examined the
monotonic relationship between the average helium difference and both the
absolute luminosity and the metallicity of the host GCs. We estimated the
statistical correlation between each pair of variables by using the Spearman's
rank correlation coefficient, $\rho$, and associated to each value of $\rho$, an
uncertainty $\sigma_{\rho}$, that was determined as in \citet{Milone14} and is
indicative of the robustness of the correlation coefficient. Shortly, we
generated 1,000 equal-size re-samples of the original dataset by randomly
sampling with replacement from the observed dataset. For each $i$-th re-sample,
we have determined $\rho_{i}$ and assumed $\sigma_{\rho}$ as the 68.27th
percentile of the $\rho_{i}$ measurements. As illustrated in
Figure~\ref{fig:corr_2}, we did not find any significant correlation between
$\rm \langle \Delta Y \rangle$ and both \feh\ ($\rho=-0.07\pm0.28$) and $\rm
M_V$ ($\rho=-0.30\pm0.24$).
It should be noted that the lack of correlation is not in contrast with
the results obtained in \citetalias{Milone17} because the average helium
abundance differences estimated in this work represent a lower limit for
the maximum helium variation in the analysed clusters.

\section{Summary and conclusions}\label{sec:end}   
Recent studies based on multiwavelength photometry of GCs have revealed that all
the analysed clusters host two discrete groups of RGB stars that correspond to
the first and the second stellar generation \citepalias{Piotto15,Milone17}.  In
this paper, we used the photometric catalogues from the \hst\ UV Legacy Survey of
Galactic Globular Clusters \citepalias{Piotto15} and the ACS Survey of Galactic
Globular Clusters Treasury Program \cite{Anderson08}, to search the RGBB of 1G
and 2G stars in a large sample of 56 GCs.

We identified, for the first time, the RGBB of both 1G and 2G stars with high
significance in 26 GCs by analysing the LF for the RGB stars of each population.
For each cluster, we estimated the location of the two RGBBs in the F275W,
F336W, F438W, F606W and F814W bands and calculated the magnitude difference
between the RGBB of 2G and 1G stars, $\rm \Delta m^{(2G,1G)}_{X}$.  When plotted
against the central wavelength of the corresponding filter, X, the quantity $\rm
\Delta m^{(2G,1G)}_{X}$ exhibits similar trends in all the analysed GCs.
Specifically, the magnitude separation between the RGBBs of 2G and 1G stars is
nearly the same in the F438W, F606W and F814W bands where the RGBB of 2G stars
is typically brighter than that of 1G stars. The relative F336W magnitude
difference between the two bumps changes significantly from one cluster to
another. In some GCs, like \ngc6723, 2G stars have a RGBB fainter than that of
1G stars in the F336W band, while in other clusters, like \ngc6352 and \ngc6752,
the two RGBBs have nearly the same luminosity. In contrast, the RGBB of 2G stars
exhibits a F275W magnitude brighter than that of 1G stars in most GCs.     

To understand the physical reasons responsible for the observed magnitude
difference of the RGBB stars we computed synthetic stellar atmospheres for RGBB
stars by assuming the chemical composition mixtures typical of 1G and 2G stars.
We compared the $\rm \Delta m_{X}^{(2G,1G)}$ values with theoretical magnitude
differences derived from the isochrones of the BaSTI databases and from synthetic
spectra. We found that the luminosity of the RGBB in the F275W and F336W filters
is strongly affected by the abundance of O and N respectively, through the
effect of the OH and NH molecular bands on the stellar atmosphere. The F438W
band is affected by strong CN and CH bands but in this case the effect on the
magnitude is significantly smaller than in the UV and never exceeds 0.015 mag.
Light elements also affect the stellar luminosity of RGBB stars in the F606W and
F814W bands, but the corresponding magnitude variation is very small, exceeding
$\sim 0.01$ mag only in GCs more metal-rich than $\feh \gtrsim -1.0$.  

Nevertheless, C-, N-, O-abundance variations alone are not able to reproduce the
observations and some helium difference between 2G and 1G stars is needed to
reproduce the observed values of $\rm \Delta m_{X}^{(2G,1G)}$. By comparing the
theoretical F606W and F814W magnitudes of 1G and 2G RGBB stars derived from
synthetic spectra and from isochrones, we estimated the average helium
difference, $\rm \langle \Delta Y \rangle$, between 2G and 1G stars in 17 GCs
with $\feh \lesssim -1.0$ and in \ngc104, for which accurate C, N, and O
abundances are available from high-resolution spectroscopy. This is the first
determination of relative helium abundance of multiple populations in a large
sample of GCs. We found that the 2G stars are more helium-rich than the 1G stars
in most GCs, and that in all the GCs the average helium difference is smaller
than $\rm \langle \Delta Y \rangle \sim 0.035$. On average, the 2G stars are
enhanced in helium by $\rm \Delta Y=0.011\pm0.002$ with respect to the 1G stars.

It should be noted that the estimated $\rm \langle \Delta Y \rangle$ are
determined through the luminosity of the RGBB and therefore are associated to
the entire stellar structure rather than to atmospheric helium abundance
variations.

The findings that the stellar populations of some GCs exhibits large helium
differences up to $\rm \Delta Y \sim 0.14$
\citep[e.g.][]{Norris04,DAntona05,Piotto07,King12} are not in contrast with the
conclusions of this paper. Indeed, both the 2G and 1G stars of the studied GCs
host sub-populations of stars with different helium and light element abundance
\citepalias[e.g.][]{Milone15b,Milone17}. For this reason, the difference between
the average helium abundance of the 2G and 1G stars is significantly smaller
than the maximum helium internal variation within each GC. 

The results of this paper, which are based on the luminosity of
the RGBBs, further corroborate similar findings based on independent techniques
and demonstrate that 2G stars are enhanced in helium, as earlier suggested by
\citet{DAntona02} on the basis of the HB morphology of some GCs.

Most scenarios on the formation of multiple populations in GCs have suggested
that 2G stars born from the material polluted from massive 1G stars. The nature
of the polluters is still debated and AGB stars, fast-rotating massive stars,
and supermassive stars are considered possible candidates
\citepalias{Renzini15}. In this context, several authors have estimated the
helium abundance that we would expect if 2G stars formed from pure ejecta coming
from a previous generation of polluting stars
\citep[e.g.][]{Ventura09,Decressin07,Denissenkov14}. 

They concluded that, if AGB or super-AGB stars are responsible for the chemical
composition of 2G stars the helium content of 2G stars would never go beyond
$\rm Y=0.40$, while in the case of fast-rotating massive stars we
would expect that some 2G stars have helium content larger, or even much larger
than 0.40 in mass fraction \citep[e.g.][]{Chantereau16}.  However caution is
necessary when using the average helium difference between 2G and 1G stars to
constrain the \textit{maximum} helium (and light elements) variations predicted
by pollution models.  Indeed, the average helium difference between 2G and 1G
stars does not necessarily reflect or correlate with the maximum internal
variations of the same element in GCs.

Both the average and the maximum helium abundance variations represent essential
ingredients to shed light on the knowledge of the formation process of 2G stars
in GCs.

\section*{Acknowledgements}
APM and AFM acknowledge support by the Australian Research Council through
Discovery Early Career Researcher Awards DE150101816 and DE160100851. EPL and APM
acknowledges support by the project ERC-StG 2016 716082 funded by the European
Research Council. SC, FD, GP, and AR acknowledge financial support from
PRIN-INAF2014 (PI S.  Cassisi).

\bsp	
\label{lastpage}

\begin{thebibliography}{99}
\bibitem[Anderson \& King(2003)]{Anderson03} Anderson, J., \& King, I.~R.\ 2003, \aj, 126, 772
\bibitem[Anderson \& King(2006)]{AndKing06} Anderson, J., \& King, I.~R.\ 2006, Instrument Science Report ACS 2006-01, 34 pages
\bibitem[Anderson et al.(2006)]{Anderson06} Anderson, J., Bedin, L.~R., Piotto, G., Yadav, R.~S., \& Bellini, A.\ 2006, \aap, 454, 1029
\bibitem[Anderson et al.(2008)]{Anderson08} Anderson, J., Sarajedini, A., Bedin, L.~R., et al.\ 2008, \aj, 135, 2055
\bibitem[Anderson \& Bedin(2010)]{Anderson10} Anderson, J., \& Bedin, L.~R.\ 2010, \pasp, 122, 1035

\bibitem[Bedin et al.(2004)]{Bedin04} Bedin, L.~R., Piotto, G., Anderson, J., et al.\ 2004, \apjl, 605, L125 
\bibitem[Bedin et al.(2005)]{Bedin05} Bedin, L.~R., Cassisi, S., Castelli, F., et al.\ 2005, \mnras, 357, 1038

\bibitem[Bellini et al.(2011)]{Bellini11} Bellini, A., Anderson, J., \& Bedin, L.~R.\ 2011, \pasp, 123, 622
\bibitem[Bellini et al.(2013)]{Bellini13} Bellini, A., Piotto, G., Milone, A.~P., et al.\ 2013, \apj, 765, 32 
\bibitem[Bellini et al.(2017)]{Bellini17} Bellini, A., Anderson, J., Bedin, L.~R., et al.\ 2017, \apj, 842, 6 

\bibitem[Bono et al.(2001)]{Bono01} Bono, G., Cassisi, S., Zoccali, M., \& Piotto, G.\ 2001, \apjl, 546, L109 

\bibitem[Bragaglia et al.(2010)]{Bragaglia10} Bragaglia, A., Carretta, E., Gratton, R., et al.\ 2010, \aap, 519, A60

\bibitem[Brown et al.(2016)]{Brown16} Brown, T.~M., Cassisi, S., D'Antona, F., et al.\ 2016, \apj, 822, 44

\bibitem[Carretta et al.(2009)]{Carretta09} Carretta, E., Bragaglia, A., Gratton, R.~G., et al.\ 2009, \aap, 505, 117
\bibitem[Carretta et al.(2010)]{Carretta10} Carretta, E., Gratton, R.~G., Lucatello, S., et al.\ 2010, \apjl, 722, L1

\bibitem[Cassisi \& Salaris(1997)]{Cassisi97} Cassisi, S., \& Salaris, M.\ 1997, \mnras, 285, 593 
\bibitem[Cassisi et al.(2016)]{Cassisi16} Cassisi, S., Salaris, M., \& Pietrinferni, A.\ 2016, \aap, 585, A124

\bibitem[Castelli \& Kurucz(2004)]{Castelli04} Castelli, F., \& Kurucz, R.~L.\ 2004, arXiv:astro-ph/0405087

\bibitem[Chantereau et al.(2016)]{Chantereau16} Chantereau, W., Charbonnel, C., \& Meynet, G.\ 2016, \aap, 592, A111

\bibitem[D'Antona et al.(2002)]{DAntona02} D'Antona, F., Caloi, V., Montalb{\'a}n, J., Ventura, P., \& Gratton, R.\ 2002, \aap, 395, 69
\bibitem[D'Antona et al.(2005)]{DAntona05} D'Antona, F., Bellazzini, M., Caloi, V., et al.\ 2005, \apj, 631, 868 
\bibitem[D'Antona et al.(2016)]{DAntona16} D'Antona, F., Vesperini, E., D'Ercole, A., et al.\ 2016, \mnras, 458, 2122

\bibitem[Decressin et al.(2007)]{Decressin07} Decressin, T., Meynet, G., Charbonnel, C., Prantzos, N., \& Ekstr{\"o}m, S.\ 2007, \aap, 464, 1029

\bibitem[de Mink et al.(2009)]{deMink09} de Mink, S.~E., Pols, O.~R., Langer, N., \& Izzard, R.~G.\ 2009, \aap, 507, L1

\bibitem[Denissenkov \& Hartwick(2014)]{Denissenkov14} Denissenkov, P.~A., \& Hartwick, F.~D.~A.\ 2014, \mnras, 437, L21

\bibitem[D'Ercole et al.(2010)]{Dercole10} D'Ercole, A., D'Antona, F., Ventura, P., Vesperini, E., \& McMillan, S.~L.~W.\ 2010, \mnras, 407, 854

\bibitem[di Criscienzo et al.(2011)]{diCriscienzo11} di Criscienzo, M., D'Antona, F., Milone, A.~P., et al.\ 2011, \mnras, 414, 3381

\bibitem[Dotter et al.(2010)]{Dotter10} Dotter, A., Sarajedini, A., Anderson, J., et al.\ 2010, \apj, 708, 698 

\bibitem[Dupree et al.(2011)]{Dupree11} Dupree, A.~K., Strader, J., \& Smith, G.~H.\ 2011, \apj, 728, 155 

\bibitem[Gilliland(2004)]{Gilliland04} Gilliland, R.~L.\ 2004, Instrument Science Report ACS 2004-01, 18 pages

\bibitem[Gratton et al.(2012)]{Gratton12} Gratton, R.~G., Carretta, E., \& Bragaglia, A.\ 2012, \aapr, 20, 50

\bibitem[Grundahl et al.(1999)]{Grundahl99} Grundahl, F., Catelan, M., Landsman, W.~B., Stetson, P.~B., \& Andersen, M.~I.\ 1999, \apj, 524, 242

\bibitem[Han et al.(2009)]{Han09} Han, S.-I., Lee, Y.-W., Joo, S.-J., et al.\ 2009, \apjl, 707, L190

\bibitem[Harris(1996)]{Harris96} Harris, W.E.\ 1996, \aj, 112, 1487

\bibitem[Iben(1968)]{Iben68} Iben, I.\ 1968, \nat, 220, 143 

\bibitem[Johnson et al.(2009)]{Johnson09} Johnson, C.~I., Pilachowski, C.~A., Michael Rich, R., \& Fulbright, J.~P.\ 2009, \apj, 698, 2048
\bibitem[Johnson et al.(2015)]{Johnson15} Johnson, C.~I., Rich, R.~M., Pilachowski, C.~A., et al.\ 2015, \aj, 150, 63 
\bibitem[Johnson et al.(2017)]{Johnson17} Johnson, C.~I., Caldwell, N., Rich, R.~M., et al.\ 2017, \apj, 836, 168 


\bibitem[King et al.(2012)]{King12} King, I.~R., Bedin, L.~R., Cassisi, S., et al.\ 2012, \aj, 144, 5

\bibitem[Kurucz(2005)]{Kurucz05} Kurucz, R.~L.\ 2005, Memorie della Societa Astronomica Italiana Supplementi, 8, 14

\bibitem[Marino et al.(2009)]{Marino09} Marino, A.~F., Milone, A.~P., Piotto, G., et al.\ 2009, \aap, 505, 1099 
\bibitem[Marino et al.(2011)]{Marino11} Marino, A.~F., Milone, A.~P., Piotto, G., et al.\ 2011, \apj, 731, 64
\bibitem[Marino et al.(2012)]{Marino12} Marino, A.~F., Milone, A.~P., Piotto, G., et al.\ 2012, \apj, 746, 14
\bibitem[Marino et al.(2014)]{Marino14} Marino, A.~F., Milone, A.~P., Przybilla, N., et al.\ 2014, \mnras, 437, 1609 
\bibitem[Marino et al.(2015)]{Marino15} Marino, A.~F., Milone, A.~P., Karakas, A.~I., et al.\ 2015, \mnras, 450, 815
\bibitem[Marino et al.(2016)]{Marino16} Marino, A.~F., Milone, A.~P., Casagrande, L., et al.\ 2016, \mnras, 459, 610

\bibitem[Milone et al.(2009)]{Milone09} Milone, A.~P., Bedin, L.~R., Piotto, G., \& Anderson, J.\ 2009, \aap, 497, 755
\bibitem[Milone et al.(2012a)]{Milone12a} Milone, A.~P., Piotto, G., Bedin, L.~R., et al.\ 2012, \aap, 537, A77
\bibitem[Milone et al.(2012b)]{Milone12b} Milone, A.~P., Piotto, G., Bedin, L.~R., et al.\ 2012, \apj, 744, 58 
\bibitem[Milone et al.(2012c)]{Milone12c} Milone, A.~P., Marino, A.~F., Cassisi, S., et al.\ 2012, \apjl, 754, L34
\bibitem[Milone et al.(2013)]{Milone13} Milone, A.~P., Marino, A.~F., Piotto, G., et al.\ 2013, \apj, 767, 120
\bibitem[Milone et al.(2014)]{Milone14} Milone, A.~P., Marino, A.~F., Dotter, A., et al.\ 2014, \apj, 785, 21
\bibitem[Milone et al.(2015a)]{Milone15a} Milone, A.~P., Marino, A.~F., Piotto, G., et al.\ 2015, \mnras, 447, 927
\bibitem[Milone et al.(2015b)]{Milone15b} Milone, A.~P., Marino, A.~F., Piotto, G., et al.\ 2015, \apj, 808, 51
\bibitem[Milone et al.(2017)]{Milone17} Milone, A.~P., Piotto, G., Renzini, A., et al.\ 2017, \mnras, 464, 3636 

\bibitem[Nardiello et al.(2015)]{Nardiello15} Nardiello, D., Piotto, G., Milone, A.~P., et al.\ 2015, \mnras, 451, 312

\bibitem[Nataf et al.(2011)]{Nataf11} Nataf, D.~M., Gould, A., Pinsonneault, M.~H., \& Stetson, P.~B.\ 2011, \apj, 736, 94
\bibitem[Nataf(2014)]{Nataf14} Nataf, D.~M.\ 2014, \mnras, 445, 3839

\bibitem[Norris(2004)]{Norris04} Norris, J.~E.\ 2004, \apjl, 612, L25

\bibitem[Pasquini et al.(2011)]{Pasquini11} Pasquini, L., Mauas, P., K{\"a}ufl, H.~U., \& Cacciari, C.\ 2011, \aap, 531, A35

\bibitem[Pietrinferni et al.(2004)]{Pietrinferni04} Pietrinferni, A., Cassisi, S., Salaris, M., \& Castelli, F.\ 2004, \apj, 612, 168 
\bibitem[Pietrinferni et al.(2006)]{Pietrinferni06} Pietrinferni, A., Cassisi, S., Salaris, M., \& Castelli, F.\ 2006, \apj, 642, 797
\bibitem[Pietrinferni et al.(2009)]{Pietrinferni09} Pietrinferni, A., Cassisi, S., Salaris, M., Percival, S., \& Ferguson, J.~W.\ 2009, \apj, 697, 275

\bibitem[Piotto et al.(2005)]{Piotto05} Piotto, G., Villanova, S., Bedin, L.~R., et al.\ 2005, \apj, 621, 777
\bibitem[Piotto et al.(2007)]{Piotto07} Piotto, G., Bedin, L.~R., Anderson, J., et al.\ 2007, \apjl, 661, L53
\bibitem[Piotto et al.(2012)]{Piotto12} Piotto, G., Milone, A.~P., Anderson, J., et al.\ 2012, \apj, 760, 39
\bibitem[Piotto et al.(2015)]{Piotto15} Piotto, G., Milone, A.~P., Bedin, L.~R., et al.\ 2015, \aj, 149, 91

\bibitem[Prantzos \& Charbonnel(2006)]{Prantzos06} Prantzos, N., \& Charbonnel, C.\ 2006, \aap, 458, 135 

\bibitem[Renzini et al.(2015)]{Renzini15} Renzini, A., D'Antona, F., Cassisi, S., et al.\ 2015, \mnras, 454, 4197

\bibitem[Sbordone et al.(2007)]{Sbordone07} Sbordone, L., Bonifacio, P., \& Castelli, F.\ 2007, Convection in Astrophysics, 239, 71

\bibitem[Silverman(1986)]{Silver86} Silverman, B.~W.\ 1986, Monographs on Statistics and Applied Probability, London: Chapman and Hall, 1986

\bibitem[Sweigart et al.(1990)]{Sweigart90} Sweigart, A.~V., Greggio, L., \& Renzini, A.\ 1990, \apj, 364, 527

\bibitem[Thomas(1967)]{Thomas67} Thomas, H.-C.\ 1967, \zap, 67, 420

\bibitem[Ventura \& D'Antona(2009)]{Ventura09} Ventura, P., \& D'Antona, F.\ 2009, \aap, 499, 835 

\bibitem[Villanova et al.(2007)]{Villanova07} Villanova, S., Piotto, G., King, I.~R., et al.\ 2007, \apj, 663, 296
\bibitem[Villanova et al.(2009)]{Villanova09} Villanova, S., Piotto, G., \& Gratton, R.~G.\ 2009, \aap, 499, 755

\bibitem[Yong et al.(2008)]{Yong08} Yong, D., Grundahl, F., Johnson, J.~A., \& Asplund, M.\ 2008, \apj, 684, 1159-1169
\bibitem[Yong et al.(2009)]{Yong09} Yong, D., Grundahl, F., D'Antona, F., et al.\ 2009, \apjl, 695, L62
\bibitem[Yong et al.(2014)]{Yong14} Yong, D., Roederer, I.~U., Grundahl, F., et al.\ 2014, \mnras, 441, 3396

\end{thebibliography}
\end{document}